\input harvmac
\input epsf

\font\myrm=cmr10 scaled 1060


\def\P#1#2{P'_{#1}(\Phi_{#2})}
\def\id{\relax{\rm {\myrm 1}\kern-.25em I}}
\def\s{\sigma}
\def\st{\tilde{\sigma}}
\def\sh{\hat{\sigma}}
\def\e{\epsilon}
\def\uh{{\hat u}}

\Title
 {\vbox{
 \baselineskip12pt
 \hbox{HUTP-01/A038}
\hbox{ILL-(TH)-01-7}
\hbox{OSU-M-2001-4}
 \hbox{hep-th/0108120}\hbox{}\hbox{}
}}
 {\vbox{
 \centerline{Geometric Transitions
and ${\cal N}=1$ Quiver Theories}
 }}
 \medskip

\centerline{F. Cachazo$^1$, S. Katz$^{2,3}$ and
C. Vafa$^1$}
\medskip
\centerline{$^1$ Jefferson Physical Laboratory}
\centerline{Harvard University}
\centerline{Cambridge, MA 02138, USA}
\medskip
\centerline{$^2$ Department of Mathematics}
\centerline{Oklahoma State University}
\centerline{Stillwater, OK 74078 USA}
\medskip
\centerline{$^3$ Departments of Mathematics and Physics}
\centerline{University of Illinois at Urbana-Champaign}
\centerline{Urbana, IL 61801 USA}
\bigskip

 \vskip .3in \centerline{\bf Abstract}
We construct ${\cal N}=1$ supersymmetric theories
on worldvolumes of $D5$ branes wrapped around 2-cycles of
threefolds which are A-D-E fibrations over a plane.
We propose large $N$ duals as geometric transitions involving blowdowns
of two cycles and blowups of three-cycles.  This yields exact predictions
for a large class of ${\cal N}=1$ supersymmetric gauge systems including
$U(N)$ gauge theories with two adjoint matter fields deformed by
superpotential terms, which arise in A-D-E fibered geometries
with non-trivial monodromies.
 \smallskip
\Date{August 2001}


\lref\gab{P. Gabriel, ``Unzerlegbare Darstellungen I (Oberwolfach
1970). Manuscr. Math {\bf 6} (1972), 71-103}
\lref\fibra{M. Bershadsky, V. Sadov, and C. Vafa, ``D-Strings on D-Manifolds,''
hep-th/9510225, Nucl.Phys. {\bf B463} (1996) 398-414 \semi S. Katz, D.R.
Morrison and M.R. Plesser, ``Enhanced Gauge Symmetry in Type II String Theory,''
hep-th/9601108, Nucl. Phys. {\bf B477} (1996) 105-140 \semi A. Klemm, W. Lerche,
P. Mayr, C. Vafa, and N. Warner, ``Self-Dual Strings and N=2 Supersymmetric
Field Theory,'' hep-th/9604034, Nucl.Phys. {\bf B477} (1996) 746-766 \semi S. Katz and C. Vafa,``Matter From Geometry,''
hep-th/9606086, Nucl.Phys. {\bf B497} (1997) 146-154 \semi N. Nekrasov, S. Gubser, and S. Shatashvili, ``Generalized
Conifolds and 4d N=1 SCFT'', hep-th/9811230, JHEP {\bf 9905} (1999) 003.}
\lref\km{S. Katz and D. Morrison, ``Gorenstein Threefold Singularities with
Small Resolutions via Invariant Theory for Weyl Groups'', J. Algebraic
Geometry {\bf 1} (1992) 449-530.}
\lref\min{M. Aganagic, private communication.}
\lref\civ{F. Cachazo, K. Intriligator, and C. Vafa, ``A Large N Duality via
a Geometric Transition'', hep-th/0103067, Nucl.Phys. {\bf B603} (2001) 3-41.}
\lref\gova{R. Gopakumar and C. Vafa, ``On the Gauge Theory/Geometry
Correspondence'', hep-th/9811131, Adv.Theor.Math.Phys. {\bf 3} (1999) 1415-1443.}
\lref\kls{I. Klebanov and M. Strassler, ``Supergravity and a Confining Gauge Theory: Duality Cascades and
$\chi$SB-Resolution of Naked Singularities'', hep-th/0007191, JHEP {\bf 0008} (2000) 052.}
\lref\nm{C. Nunez and J. Maldacena, ``Towards the large N limit of pure N=1
super Yang Mills'', hep-th/0008001, Phys.Rev.Lett. {\bf 86} (2001) 588-591.}
\lref\va{C. Vafa, ``Superstrings and Topological Strings at Large N'',
hep-th/0008142.}
\lref\mdo{M. Douglas,``Enhanced Gauge Symmetry in M(atrix) Theory'',
hep-th/9612126, JHEP {\bf 9707} (1997) 004.}
\lref\dmo{M. Douglas and G. Moore, ``D-branes, Quivers, and ALE Instantons'',
hep-th/9603167.}
\lref\doug{E. Diaconescu, M. Douglas and J. Gomis, ``Fractional Branes and Wrapped
Branes'', hep-th/9712230, JHEP {\bf 9802} (1998) 013.}
\lref\bkl{J. Bryan, S. Katz, and N.C. Leung, ``Multiple covers and the integrality conjecture for rational curves in Calabi-Yau
    threefolds'', math.AG/9911056.}
\lref\small{S. Katz, ``Small resolutions of Gorenstein threefold
singularities'', Cont. Math. {\bf 116} (1991)61-70.}
\lref\laufer{H. Laufer, ``On $CP^1$ as exceptional set'', Recent
Developments in Several Complex Variables, Ann. of Math. Stud. {\bf 100} (1981), 261-276.}
\lref\klwi{I. Klebanov and E. Witten, ''Superconformal Field Theory on
Threebranes at a Calabi-Yau Singularity'', hep-th/9807080, Nucl.Phys. {\bf B536} (1998) 199-218.}
\lref\neg{N. Nekrasov, S. Gubser, and S. Shatashvili, ``Generalized
Conifolds and 4d N=1 SCFT'', hep-th/9811230, JHEP {\bf 9905} (1999) 003.}
\lref\gvw{S. Gukov, C. Vafa, and E. Witten, ``CFT's From Calabi-Yau
Four-folds'', hep-th/9906070, Nucl.Phys. {\bf B584} (2000) 69-108.}
\lref\shapv{A. Shapere and C. Vafa, ``BPS Structure of Argyres-Douglas
Superconformal Theories'', hep-th/9910182.}
\lref\tv{T.R. Taylor and C. Vafa, ``RR Flux on Calabi-Yau and Partial
Supersymmetry Breaking'', hep-th/9912152, Phys.Lett. {\bf B474} (2000) 130-137.}
\lref\cfikv{F. Cachazo, B. Fiol, K. Intriligator, S. Katz, and C. Vafa,
in preparation.}
\lref\kkal{S. Katz, S. Kachru, A. Lawrence, and J. McGreevy, ``Open string
instantons and superpotentials'', hep-th/9912151, Phys.Rev. {\bf D62} (2000)
026001; ``Mirror symmetry for open strings,'' hep-th/0006047,
Phys.Rev. {\bf D62}
(2000) 126005.}
\lref\kron{P.B. Kronheimer, ``The construction of ALE spaces as hyperKahler
quotients,'' J. Differ. Geom. {\bf 28} (1989) 665.}
\lref\kac{V.G. Kac, ``Infinite root systems, representations of graphs and
invariant theory,'' Inv. Math. {\bf 56} (1980) 263.}
\lref\wit{E. Witten, ``Branes And The Dynamics Of QCD,'' hep-th/9706109,
Nucl.Phys. {\bf B507} (1997) 658-690. }
\lref\av{M. Aganagic and C. Vafa, ``Mirror Symmetry, D-Branes and Counting
Holomorphic Discs,'' hep-th/0012041}
\lref\potential{S. Katz, `` Versal deformations and superpotentials for
rational curves in smooth threefolds,'' math.AG/0010289}
\lref\mayr{P. Mayr, ``On Supersymmetry Breaking in String Theory and its
Realization in Brane Worlds,'' hep-th/0003198, Nucl.Phys. {\bf B593} (2001) 99-126}
\lref\morec{J. Edelstein, K. Oh, and
R. Tatar, ``Orientifold, Geometric Transition and Large N Duality for SO/Sp
Gauge Theories,'' hep-th/0104037, JHEP {\bf 0105} (2001) 009 \semi
K. Dasgupta, K. Oh, and R. Tatar, ``Geometric Transition, Large N Dualities
and MQCD Dynamics,'' hep-th/0105066 \semi
S.B. Giddings, S. Kachru and J. Polchinski, ``Hierarchies From Fluxes in
String Compactifications,'' hep-th/0105097 \semi
K. Dasgupta, K. Oh, and R. Tatar,
``Open/Closed String Dualities and Seiberg Duality from Geometric Transitions in
    M-theory,'' hep-th/0106040.}

\newsec{Introduction}
Geometric transitions in the context of D-branes have
been shown to be related to large $N$ dualities.  This
has been seen in the context of topological strings
\gova\ and more recently in the context of
superstrings \kls\nm\va\civ . See also some recent works
in this direction \morec.  The aim
of this paper is to enlarge this class of dualities by
considering a class of ${\cal N}=1$ supersymmetric
gauge theories in the context of wrapped branes for A-D-E
fibered Calabi-Yau 3-folds.  These
can be viewed as natural generalization of the dualities
of CY 3-folds studied in \civ .
These include ${\cal N}=2$ gauge theories with gauge groups
given by the quiver diagrams of A-D-E with bi-fundamental
matter dictated by the diagram, deformed by superpotential
terms for the adjoint fields, breaking the
supersymmetry to ${\cal N}=1$.  Some examples
of this kind were studied in \civ\ and the exact
results for gauge theory were recovered more
naturally from the geometric large $N$ dual.

We also construct, by considering A-D-E fibered geometries
with monodromy,
 ${\cal N}=1$  $U(N)$ gauge theories with
two adjoint fields with certain superpotential terms.
In particular we geometrically engineer theories with superpotential
$$W=P_{p+2}(X)+P_{q+2}(Y)+P_{r+2}(X+Y)$$
 where $X,Y$ are adjoint fields
and $P_i$ denotes traces of polynomials of degree $i$ in the variable.
The $p,q,r$ denote the number of nodes of the D-E Dynkin diagrams in each
of the three disconnected components after removing the trivalent node.
 Obtaining
exact results for the vacuum structure of these theories has
been difficult with the available techniques for dealing
with supersymmetric gauge theories.  This is because theories
with two adjoint fields is neither close to being an ${\cal N}=4$
system which has 3 adjoint fields, nor close to being an ${\cal N}=2$
system which has only 1 adjoint field.
  Nevertheless geometric duals we propose provide exact information
for the vacuum structure of these theories with two adjoint fields.

The organization of this paper is as follows:  In section 2, we review
A-D-E singularities and their deformations.  We also review
the worldvolume theory of branes wrapped
around cycles of these geometries.  In section
3 we consider 3-fold geometries obtained by fibering
the A-D-E singularities over a plane.  The corresponding wrapped
branes yield ${\cal N}=1$ quiver theories.
  In section 4 and 5 we study
the Higgs branches of these theories.  In section 6 we consider
the quiver theories in case the fibration involves non-trivial
Weyl group monodromies.  In section 7 we present detailed
examples of the monodromic cases, involving ${\cal N}=1$ gauge theories
with one or two adjoint fields.
 Finally in section 8 we propose the corresponding
large $N$ dual geometries.

\newsec{A-D-E singularities in dimension 2}

Consider ALE spaces with A-D-E singularities at the origin. These spaces
are constructed as the quotient of ${\bf C}^2$ by a discrete subgroup of $SU(2)$,
$\Gamma$. The correspondence between the
groups $\Gamma$ and the geometry is $cyclic \leftrightarrow A$, $dihedral
\leftrightarrow D$ and $exceptional \leftrightarrow E$.  As is well
known these geometries are singular and can be viewed as the
hypersuface
 $f(x,y,z)=0$ of ${\bf C}^3$:
$$
\eqalign{
A_r:& \qquad f=x^2+y^2+z^{r+1}\cr
D_r:& \qquad f=z^{r-1}+zy^2+x^2\cr
E_6:& \qquad f=y^3+z^4+x^2\cr
E_7:& \qquad f=y^3+yz^3+x^2\cr
E_8:& \qquad f=y^3+z^5+x^2}
$$
Note that these spaces are singular at the origin $x=y=z=0$.
There are two ways to desingularize these spaces.  One way is by
deforming the defining equation $f=0$ by adding relevant deformations.
The other way is by blowing up the singularity.

Consider deforming these by relevant deformations, so that
$f=df=0$ has no solutions.  Then the resulting
space is non-singular.  Let $r$ denote the rank
of the corresponding A-D-E.  There are $r$ such deformations
for each singularity type. It is convenient, as will be
explained below, to introduce coordinates
in the
deformation space $t_i$ \km\
with $i=1,\ldots ,r$ for $A_{r-1}$, $D_r$ and $E_r$. For $A_{r-1}$ one has
to impose one constraint. The deformed
equations $f(x,y,x,t_i)$ are given by,
\eqn\Adef{A_{r-1}  \quad : \quad x^2+y^2+\prod_{i=1}^{r}(z+t_i)  \qquad
\sum_{i=1}^{r}t_i = 0}
\eqn\Ddef{D_{r}  \quad : \quad x^2+y^2z
+{\prod_{i=1}^{r}(z+t_i^2)-\prod_{i=1}^rt_i^2 \over z}+2 \prod_{i=1}^r t_i
y }
$$E_6 \quad : \quad x^2+z^4+y^3+\epsilon_2 yz^2+\epsilon_5 yz+\epsilon_6
z^2+\epsilon_8 y+\epsilon_9 z+\epsilon_{12}$$
$$E_7 \quad : \quad
x^2+y^3+yz^3+\epsilon_2y^2z+\epsilon_6y^2+\epsilon_8yz+\epsilon_{10}z^2+\epsilon_{12}y+\epsilon_{14}z+\epsilon_{18}$$
$$E_8 \quad : \quad x^2+y^3+z^5+\epsilon_2yz^3+\epsilon_8yz^2+
\epsilon_{12}z^3+\epsilon_{14}yz+\epsilon_{18}z^2+\epsilon_{20}y+\epsilon_{24}z
+\epsilon_{30}$$
where $\epsilon_i$ are complicated homogeneous polynomials \foot{The
explicit form of these for $E_6$ and $E_7$ can be found in appendix 1 and 2
of \km\ respectively.} of $t_j$'s of
degree $i$ and invariant under the permutation of $t_j$'s.
The importance of the choice of
canonical coordinates $t_i$ is that roughly speaking
they measure the holomorphic volume of $S^2$'s in the geometry:
Upon a generic such deformation
the space $f=0$ admits $r$ non-vanishing $S^2$'s.  Moreover
a basis can be chosen so that the intersection of these $S^2$'s
is the same as that of the corresponding Dynkin diagram. In
fact there are as many deformation parameters as $S^2$'s and we can
relate them to the ``holomorphic volume'' of the corresponding 2-cycle.
This we define as the integral of the holomorphic 2-form over the
corresponding $S^2$:
$$\alpha_i=\int_{S^2_i}\omega =\int_{S^2_i}{dxdy\over z}$$
Thus the set of $\alpha_i$ will be identified with
the simple roots of the corresponding Dynkin system (Figure 1).
It is also natural to
use $\alpha_i$ as the $r$ deformation parameters for $f$ and
write $f(x,y,z;\alpha _i)$. In fact the $\alpha_i$ are very simply
related to the $t_i$.
Namely, up to a constant factor we have
\eqn\labA{A_{r}: \quad \alpha_i = t_i - t_{i+1}  \quad  i=1, \ldots ,r}
\eqn\labD{D_{r}: \quad \alpha_i = t_i - t_{i+1}  \quad  i=1, \ldots ,r-1 \quad
{\rm{and}} \quad \alpha_r = t_{r-1}+t_r}
\eqn\labE{E_r: \quad \alpha_i = t_i-t_{i+1} \quad i=1, \ldots ,r-1 \quad
{\rm{and}}
\quad \alpha_r=-t_1-t_2-t_3}
In the above parametrization the root lattice of $A_r$ and $E_r$
are realized as a hyperplane in $R^{r+1}$ and that of $D_r$ is realized
in $R^r$.  Moreover the identification
of the root vectors with vectors in this space can be
read off from the above equations.  For example for
$A_r$ they are identified with $\uh_i-\uh_{i+1}$, where $\uh_i$ denote
unit vectors of $R^{r+1}$.   Moreoever in the $A_r,D_r$ case the
canonical Euclidean inner product on $R^{r+1}$ and $R^{r}$ induce
the Cartan inner product on the roots.  In the $E_r$ case the $R^{r+1}$
has signature $(r,1)$ with $\uh_{r+1}\cdot \uh_{r+1} =-1$ and $\uh_i\cdot \uh_i =1$
for $i=1,...,r$.  In the $E_r$ case the roots are identified with
$\uh_i-\uh_{i+1}$ for $i=1,...,{r-1}$ and the r-th root is identified with
$\uh_r-\uh_1-\uh_2-\uh_3$.

The degrees of polynomials $\epsilon_i$ as a function of $t$'s
follows from the quasi-homogeneity of $f$. The choice
of $t$'s as opposed to $\alpha_i$ may appear unmotivated.
The reason that we choose
$t_i$'s as the basic variables
 has to do with the fact that the corresponding Weyl
group of the singularity has a subgroup given by the permutation
group $S_r$ which in the $t$-basis simply act as permutation of the
$t$'s (and in the $A_r$ case it is $S_{r+1}$ and we have one more
$t$ with one extra constraint as given above).  The action of the
Weyl group on the parameters turns out to be very important
for our considerations in this paper.

\bigskip
\centerline{\epsfxsize=0.55\hsize\epsfbox{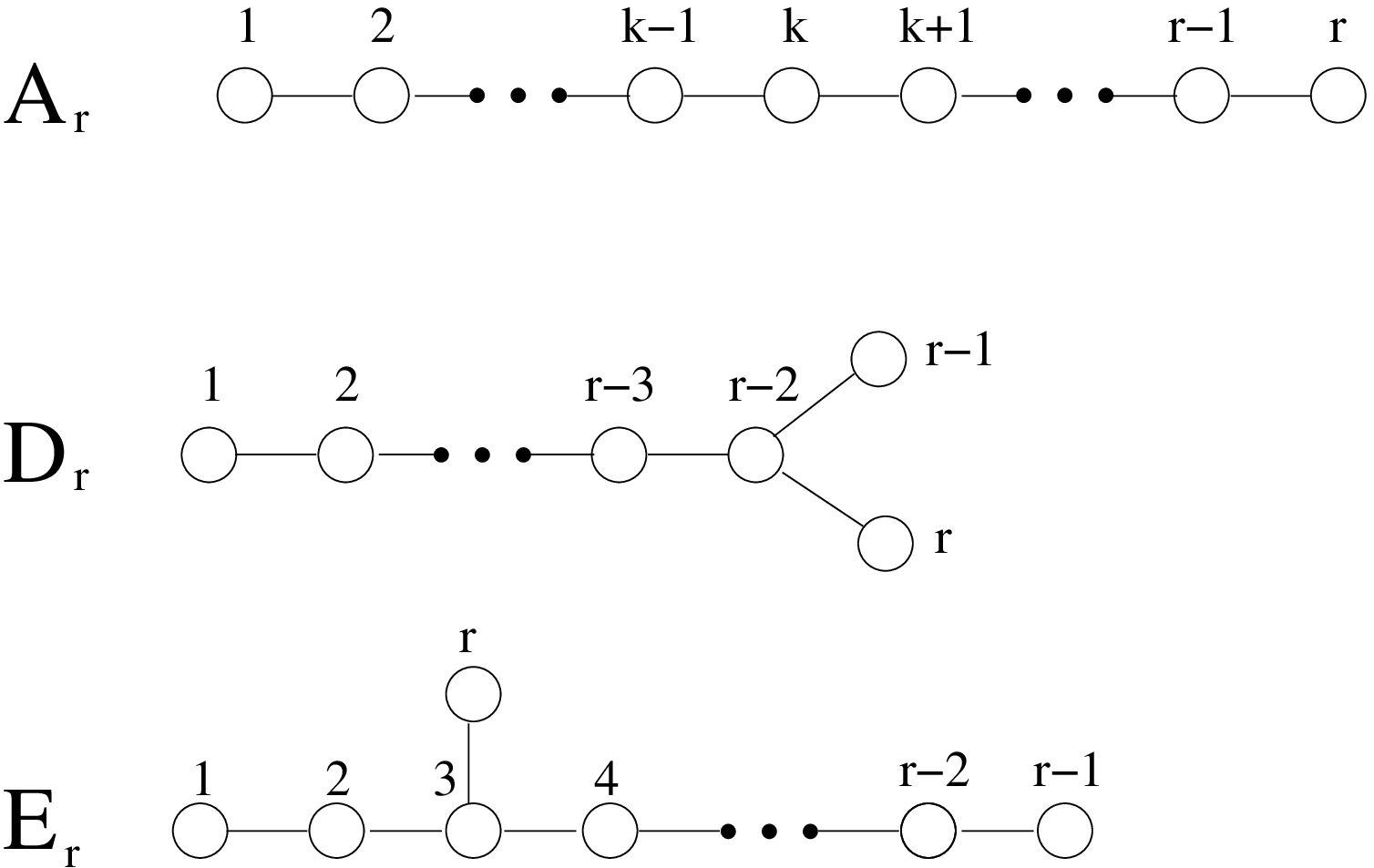}}
\noindent{\ninepoint\sl \baselineskip=8pt {\bf Figure 1}:{\sl A-D-E Dynkin
diagrams together
with our convention for the labeling of the nodes
(not to be confused with Dynkin numbers which are provided
in Figure 2).}}
\bigskip

Instead of complex deformation, one can also blow up the singularity
to get an $S^2$.  Indeed one can consider both complex deformation and
blow up, which gives a three dimensional space of deformation
of the ALE metric for each $S^2$. Let us denote the corresponding real
Kahler parameter by $r_i$.   In other words
$$r_i=\int_{S^2_i} k$$
where $k$ is the Kahler form.
The metric one obtains is hyperKahler
and under the $SO(3)$ rotation the $r_i$ and $t_i$ mix.  Once
we consider fibering this geometry over a plane,
the rotation mixing them
will no longer be a symmetry.  In string theory
one can also turn on $B$ fields.  In the context of type IIB
theory that we will be considering in this paper, there are two
choices the $NS$ field $B^{NS}$ and the $R$  field $B^R$,
and we can turn on both of them.  Thus for each $S^2$ we have
altogether a 5 parameter family of deformations in type IIB string theory.
The ``stringy'' volume of the i-th $S^2$ is given by \mdo\
\eqn\quvo{V_i=({(B^{NS}_i)}^2+r_i^2+|\alpha_i|^2)^{1/2}}

\subsec{Wrapped D5 Branes}
Now consider wrapping $N_i$ D5 branes around an $S^2_i$ of the deformed ALE space,
and occupying an $R^4\subset R^6$ subspace of the non-compact spacetime.
This gives rise to an ${\cal N}=2$ supersymmetric
$U(N_i)$ gauge theory on $R^4$.
The coupling constant of this gauge group
is given by
$${1\over (g^{YM}_i)^2}={V_i\over g_s}$$
where $V_i$ is the quantum volume of the $S^2_i$ given by \quvo\
and $g_s$ denotes the string coupling constant.  The $B^R_i$ field
plays the role of the theta angle for this gauge theory.
The diagonal components of the scalar
field $\Phi_i$ in the adjoint representation of $U(N)$ correspond
to moving the branes on the $R^2\subset R^6$ transverse subspace
to the branes. Below we will parametrize this $R^2$ subspace by the
complex coordinate $t$.  The parameters $\alpha_i$ and $r_i$ form
a triplet of ${\cal N}=2$ supersymmetric FI terms \dmo,
for the gauge theory on the brane.  In an ${\cal N}=1$ superspace
formalism the $\alpha_i$ appear in the superpotential term
$$\delta W=\int d^2\theta [\alpha_i {\rm Tr} \Phi_i]$$
and the $r_i$ term is the ${\cal N}=2$ supersymmetric completion
of this term which appears as an ordinary ${\cal N}=1$ FI term.

If we try to find the critical points of $W$ we might think
there is a contradiction with $dW/d\Phi_i=0$ as that would
lead to $\alpha_i =0$.  However we know that there are supersymmetric
wrapped branes even if $\alpha_i \not=0$.  This puzzle was noted
in \doug\ where it was pointed out that since
$W$ is linear in $\Phi_i$ in this case it just trivially adds a constant
energy to the action, and we can still preserve the supersymmetry for
non-zero holomorphic 2-volume $\alpha_i$.  However if the superpotential
was not linear in $\Phi_i$, as will not be the case
when we consider fibering this geometry over the plane, the
condition for supersymmetry will require being at the critical
points of $W$.

Now consider wrapping branes about all $S^2_i$'s corresponding
to the i-th node of the Dynkin diagram.
Note that wrapping around any other $S^2$ can be viewed
as a bound state of the configuration of this basis
of $H_2$ homology of ALE. If $N_i$ denotes the number
of the corresponding branes we obtain an ${\cal N}=2$ gauge theory
system with gauge group
$$G=\prod_{i=1}^{r} U(N_i)$$
Moreover for each neighboring (i.e. intersecting) $S^2_i, S^2_j$ we get
an ${\cal N}=2$ hypermultiplet, which can be viewed in ${\cal N}=1$
superfield terminology as two chiral multiplets $Q_{ij}$ and $Q_{ji}$
which transform as $(N_i,{\overline N_j})$ and $(N_j,{\overline N_i})$
respectively. The superpotential term for this theory can be written
in the ${\cal N}=1$ superspace notation as
\eqn\suppo{W=\int d^2\theta (\sum_{i,j}s_{ij} {\rm Tr} Q_{ij} \Phi_j Q_{ji} -
\alpha_i {\rm Tr} \Phi_i)}
where $|s_{ij}|=1$ for intersecting $S^2$'s and zero otherwise.
Moreover the sign assignment on the $s_{ij}$ is such that it is an
anti-symmetric matrix $S$.  In other words if the intersection form of the Dynkin
diagram is given as $A+A^t$ where $A$ is an upper triangular matrix
with $1$ on the diagonal, $S=A-A^t$.  There are various branches
for these theories, which are obtained by considering
the critical points of $W$, leading to
\eqn\crtw{s_{ij}Q_{ij}Q_{ji}=\alpha_i \qquad Q_{ij} \Phi_j=\Phi_i Q_{ji}}
One should also consider the D-term equations.
However, if one is just interested in parameterizing the space
of solutions this latter constraint, together with gauge invariance
under $G$, is equivalent to considering the space of solutions
to \crtw\ subject to the equivalence given by the complex gauge
transformation $G^c=\prod_i GL(N_i)$ (see e.g.
\ref\luta{M.A. Luty and W. Taylor IV, ``Varieties of Vacua in
Classical Supersymmetric Gauge Theories,''
Phys.Rev. {\bf D53} (1996) 3399.}) as long
as FI terms $r_i=0$ which we will mainly assume (if $r_i\not =0$
one typically obtains a blow up of the moduli space).
The space of solutions to
\crtw\ modulo $G^c$ has been known for a long time by mathematicians
\gab\kac\ and the answer is as follows:
There is one irreducible representation of the above algebra for each
positive root $\rho^k \in \Delta ^+$ where $\Delta^+$ denotes
the set of positive roots of A-D-E.  Moreover if we expand
$$\rho^k = \sum_{i=1}^r n_i^k e^i$$
where $e^i$ denote the simple roots, then the dimension of the $i$-th
vector space is $n_i^k$ dimensional.   For each branch
there is a condition that
$$\sum_i n_i^k \alpha_i=0,$$
 which
however is more than necessary to preserve the supersymmetry for a
single brane
as was discussed before. However if we wish to use all such
branches at the same time,
if we had more than one brane, then these conditions
will generically become also necessary conditions for preserving
supersymmetry.   Also these
conditions become necessary conditions, even if
we consider a single brane, when we consider fibering
these geometries, which leads to superpotentials which are not
linear in $\Phi$.

 This result implies that our
gauge theory will have various branches given by how many copies of each
of these irreducible representations we have.  Let $M_k$ denote the
number of times the $k$-th branch appears.  Then, since the total
dimension of the $i$-th space should be given by $N_i$, then we must have
$$N_i=\sum_k M_k n_i^k.$$
Moreover this Higgses the gauge group to
$$\prod_i U(N_i) \rightarrow \prod_k U(M_k).$$

This result can be easily understood in terms of primitive 2-cycles
of the A-D-E geometry. There is one primitive 2-cycle for every positive
root of the corresponding A-D-E.  In fact, in the context
of type IIA superstring theory, wrapping D2 branes about these cycles give
rise to the degrees of freedom for the gauge multiplet
of the corresponding A-D-E gauge group.  Thus the above
classification simply implies that with a total number of branes we
can wrap them about all the primitive 2-cycles subject only to the condition
that the $H_2$ class of the internal charge of branes are preserved.

\subsec{Stringy Orbifolds and Affine A-D-E}
In the context of string theory when one studies ${\bf C}^2/ \Gamma$
one obtains $r+1$ choices for the basis of wrapped D-branes,
where $r$ is the rank of the corresponding A-D-E \dmo .  The extra
charge in this case arises from the $H_0$ class of ALE. The choice of the
basis for the cycles are now in 1-1 correspondence with the nodes
of affine A-D-E.  The classes corresponding to $H_2$ can be read
off from the usual correspondence with the affine Dynkin diagram.
In particular the $H_2$ class of
\eqn\gana{H_2(\delta = e_0+ \sum_{i=1}^r d_i e_i =\sum_{i=0}^r d_i e_i )=0}
where $d_i$ denotes the Dynkin index of the i-th node and
$d_0=1$ and $e_0$ corresponds to the extended node of the
affine Dynkin diagram.  Moreover the $H_0$ class of ALE is
generated by the class of the brane
$\delta$ given above.  In the context of type IIB string theory
that we are considering we have $r+1$ classes of wrapped
brane charges:  $r$ types correspond to  D5 branes wrapped
over two cycles in $H_2$ and $1$ type correspond to D3 brane which is
a point on the ALE and corresponds to $H_0$.
If we wrap $N_i$ branes around the $e_i$ ( which we now include
$N_0$ branes wrapped around $e_0$) , then we get the gauge theory
corresponding to the affine A-D-E quiver \dmo .  Again
we end up as in the A-D-E quiver theories with the gauge group
$\prod_{i=0}^r U(N_i)$ and a pair of chiral multiplets
$Q_{ij} $ and $Q_{ji}$ for each link of the affine Dynkin diagram,
with the superpotential given in \suppo . Note that since the
$H_2$ class of $\delta$ is zero, and $\alpha_i$ are the holomorphic
volumes of the 2-cycles, we learn that
\eqn\cons{\int_\delta \omega =\sum_{i=0}^r d_i \alpha_i =0}
Finding the Higgs branches
by finding the critical points of \suppo\ for the affine A-D-E
case leads to the existence of one irreducible representation of the
algebra for each positive root of the affine A-D-E \kac\
(see the appendix of \ref\dofi{M.R. Douglas, B. Fiol,
C. Romelsberger, ``The spectrum of BPS branes on a noncompact Calabi-Yau,''
hep-th/0003263 .}\ for a review of this result).  The roots
of affine A-D-E can
be represented
by vectors in an $r+1$ dimensional lattice which is an
extension of the root lattice of A-D-E by a one dimensional lattice,
of the form $(R,n)$ where $R$ is a vector in
the root lattice of A-D-E and $n$ is an integer.  In this
basis $e_i$ for the affine case correspond to $(e_i,0)$ for $i>0$
and $e_0=(-\rho,1)$, where $\rho =\sum_{i=1}^r d_i e_i$.
The positive
roots of affine A-D-E are of two types: ``real'' and
``imaginary''.  The real roots correspond to rigid representations
with no moduli and the imaginary roots correspond to representations
with moduli.  The real
positive roots are given by the vectors
\eqn\afbr{(\Delta,n^+)\quad {\rm and} \quad (\Delta^+,0)}
where $\Delta$ denotes the set of {\it all} roots of A-D-E
and $n^+>0$ is a positive integer. Any real positive root can be written as
$$\rho^k =\sum_{i=0}^r n_i^k e_i=n(e_0+\sum_{i=1}^r d_ie_i )+(\beta ,0)$$
for some $n_i^k\geq 0$, $n\geq 0$
and where $\beta \in \Delta$.
 The dimension of the $i$-th vector
space associated with $\rho^k$ is given by $n_i^k$.  Moreover
for this branch we would get a necessary condition that
$$\sum_{i=0}^r n_i^k{\alpha_i}=\alpha (\beta) =0$$
where $\alpha (\beta )$ denotes the holomorphic
volume of the 2-cycle associated with $\beta$.  Note again,
as discussed before, this condition which is sufficient
for supersymmetry preservation is not strictly necessary
for a single brane but is necessary for a collection of branes
or if we consider fibrations of this geometry, as we will be
interested to do later in this paper.

The ``imaginary'' positive roots of the affine A-D-E are of the form
$N\delta = N(0,1)$, with $N>0$
\kac.  Note that for these branches the dimension of the vector
space for the $i$-th gauge group is $Nd_i$, where $d_i$ is
the Dynkin number associated with the i-th node.  These correspond
to branches with moduli.  Moreover for any decomposition $N=\sum N_k$, this branch
includes in its moduli
the subbranches corresponding to $\sum N_k \delta$ .
So for any representation corresponding to imaginary roots we will only have to label the total
number $N$.

  Let $M_k$ denote the number of
times that $\rho^k$ appears in a given
Higgs branch. Let $N$ denote the total number
of times the imaginary root $\delta $ appears.  The latter
representation has a moduli
which is isomorphic to the $N$ fold symmetric product of the ALE space
\kron.
Then by brane charge conservation we have
$$N_i=\sum_k M_k n_i^k+N d_i$$
Moreover  we get the Higgsing
$$\prod_{i=0}^r U(N_i)\rightarrow U(N)\times \prod_k U(M_k)$$
(where for a generic point on the moduli of the $U(N)$ theory
it further higgses to $U(1)^N$).
Just as in the non-affine case, we can explain
the existence of these branches by the fact that there are
as many non-trivial primitive element of $H_2$ as the root
lattice $\Delta$, and they can have arbitrary number of
internal $H_0$ class bound to them.  Up to a choice of an
overall $Z_2$ reflection, having to do with choice of brane
versus anti-brane, this explains the formula \afbr\ for
the choices of the branches.  Moreover the branch corresponding
to the imaginary roots correspond to having only the $H_0$ class.
Thus if we have $N$ of them, it should have a
Higgs branch moduli corresponding to the N-fold symmetric
product of the ALE space, as expected.

\newsec{Fibration of A-D-E Spaces}

So far we have only talked about the 2-fold internal
geometry. We wish to consider
the fibration of these spaces over
a complex plane denoted by $t$
(when we include the D5 branes this is identified with the complex
dimension one space transverse to the brane).  Similar
kinds of fibrations have been considered in the physics
\fibra\ and math literature \km .
We fix the complexified Kahler class of the 2-fold
geometry and vary the complex moduli of the ALE space.
 We consider $t_i=t_i(t)$ or
equivalently, $\alpha_i=\alpha_i(t)$.
Having a well defined fibration allows $\alpha_i$
not to be single valued functions of $t$.  This is because any
global diffeomorphism, which is given by the Weyl group of the corresponding
A-D-E leaves the geometry invariant but acts on $\alpha_i$ by the
corresponding Weyl group action.
The generic fibrations will involve such monodromic actions
on $\alpha_i$.  These we will call {\it monodromic fibrations}.
One can also consider the case of fibrations where $\alpha_i(t)$ are
single valued functions of $t$.  This we will call {\it non-monodromic
fibrations}.

Consider first the $A_1$ case.
In this case we have an
equation for the local 3-fold given by
$$f=x^2+y^2+ (z+t_1)(z+t_2)=0  \quad {\rm{with}} \quad t_1+t_2 =0 $$
using that $\alpha = t_1-t_2 = 2t_1$ and a trivial rescaling of the
equation we get the familiar form, and recalling that $\alpha $
is a function of $t$ we have
\eqn\conif{f=x^2 + y^2 + z^2 + \alpha (t)^2=0.}
Over each point $t$, there is a non-trivial $S^2$ whose
holomorphic volume is given as $\int_{S^2} \omega =\alpha(t)$.
For a well defined fibered geometry we must have $\alpha^2(t)$
to be a well defined function of $t$.  However this leaves
room for $\alpha (t)$ to have branch cuts in the $t$-plane, which
would correspond to monodromic fibrations.

\subsec{Adding in the Branes}
Now consider
wrapping a D5 brane around $S^2$ fiber, just as before
and consider the total fibered geometry including the D5 brane.
  As noted
before, the modulus $t$ corresponds
to the scalar field $\langle \Phi
\rangle  =t$ on the D5 brane worldvolume, which
is in the adjoint representation of the gauge group on the
D5 brane (for $N$ D5 branes it will be an $N\times N$ matrix).
Using the relation between BPS tension of domain walls and the
value of the superpotential, the computation of the superpotential
as a function of the modulus
in this case boils down to the computation of the integral of holomorphic
3-form $\Omega=\omega \wedge dt$ over a 3-chain ending on $S^2$, which
is parametrized by $t$ (as in \wit\kkal\av ).  In other words we have
$$W(\Phi)=\int_{S^2(t)\times I}^{t=\Phi}\omega\wedge dt =\int_I^{t=\Phi}
\alpha(t)dt$$
where $I$ is an interval in the $t$-plane.  shifting the origin
of the interval shifts $W$ by an irrelevant constant.  From this
we see that
$${dW\over d\Phi} =\alpha(\Phi)$$
This in particular means that if we wish to induce a superpotential
$P(\Phi)$, we have to define
\eqn\impo{\alpha(t)={dP(t)\over dt}.}

Note that the condition that we are at the critical point
of the superpotential means that $\alpha(t)=0$, i.e., this
corresponds to holomorphically shrunk $S^2$.  Now consider blowing up
the quantum volume of $S^2$.  Note that the above
superpotential makes sense only when the
fibration is non-monodromic, i.e. the $H_2$ class where the brane
is wrapped is invariant, which in this case means that $\alpha (t)$
is a single valued function of $t$.  We will assume this is the case.
  In this way we obtain a $U(N)$ gauge theory
with ${\cal N}=1$.
Moreover the blowup mode does not affect the superpotential
(the Kahler deformation does not affect the superpotential computation
above), and so we end up with a gauge theory with superpotential
$W=P(\Phi)$.  Of course if we have $N$ such D-branes we obtain
the same superpotential for the $U(N)$ gauge
theory where we now have to insert the trace.
The quantum volume of the blown up $S^2$, for $\alpha =0$ is given by
$V=(r^2+B_{NS}^2)^{1/2})$,  which determines the
coupling
constant of the $U(N)$ gauge theory to be $1/g_{YM}^2= V/g_s$
as discussed before. The ${\cal N}=1$ FI term is given by $r$
(which in this case is superfluous as there is no field charged
under the $U(1)$).
 If we wish to have no ${\cal N}=1$
 FI term, then $r=0$ and $V=B_{NS}$.
Note that the gauge theory we have geometrically engineered
for this case was already studied in \kkal \civ .
 The observation above for the derivation
of the relation between the
3-fold geometry and the gauge theory superpotential
was also made independently by Aganagic \min.

Note that the ${\cal N} = 2$ FI term in the context of trivial fibering
discussed in the previous section is a special case of this
formula where $\alpha$ is a constant and we get
$${dW\over d \Phi} =\alpha \rightarrow W=\alpha Tr \Phi.$$

For the same $A_1$ case we could consider the affine situation.
In this case the geometry of the underlying manifold will be the
same as we have discussed.  The only difference is in the choice
of the branes we consider, i.e. we also allow D5 branes bound
to D3 branes.  Now we will have the gauge group $U(N_0)\times U(N_1)$
where $N_0,N_1$ denote the number of branes wrapped around the
classes corresponding to each one of the two nodes of the affine $A_1$
Dynkin diagram.  In addition we will have two pairs of chiral multiplets
$Q^i_{01}, Q^i_{10}$ where $i=1,2$ correspond to the two links
of $A_1$.  There will also be a superpotential of the form
$$W=\int d^2\theta \sum_i[{\rm Tr}Q^i_{01}\Phi_1 Q^i_{10}-{\rm Tr}Q^i_{10}
\Phi_0 Q^i_{01}]+W_0(\Phi_0)+W_1(\Phi_1)$$
where we can identify $W_i(\Phi_i)=P_i(\Phi_i)$ where $P_i'(t)=\alpha_i (t)$.
Moreover using the condition that $\sum_{i=0}^r d_i\alpha_i=0$ we learn
that $W_1(\Phi_1)=P_1(\Phi_1)$ and $W_0(\Phi_0)=-P_1(\Phi_0)$.
The special case where $\alpha (t)=mt$ corresponding to addition
of mass terms to the adjoint fields
 was studied in \klwi\ and further elaborated in \neg\ from the
point of view we are pursuing here.

So far we have presented the construction for the case of $A_1$ and
affine $A_1$.
The extension of this construction to general
A-D-E theories with superpotential terms is straightforward.  We consider
non-monodromic fibrations so that we can wrap branes about non-trivial
2-cycles in the fibered geometry.  In this case $t_i(t)$ or equivalently
$\alpha_i (t)$ are single valued functions of $t$ and we take each one
to be a polynomial in $t$.
 Moreover
we define $\alpha_i(t)=dP_i(t)/dt$, leading
to a superpotential term
\eqn\superpotential{W=\int d^2\theta \sum_{ij}s_{ij}{\rm Tr}Q_{ij}\Phi_j Q_{ji}+ \sum_{i}{\rm
Tr} P_i(\Phi_i).}
This formula is equally valid
for the affine as well as the non-affine case.  In the affine
case one will have a constraint from $\sum_i d_i \alpha_i=0$
which gives rise to
$$\sum_{i=0}^r d_i P_i(t) =0$$
as a functional equation satisfied by the superpotentials $P_i(\Phi_i)$.
This class of theories we call ${\cal N}=1$ A-D-E quiver theories,
and we shall now turn to the study of their Higgs branch.   The affine
A-D-E case with mass terms, where $P_i'(t)=m_i t$ has already been
geometrically engineered in exactly this way in \neg\ for the case
where $N_i=N d_i$ (and for the affine $A_1$ generalized
to arbitrary ranks with quadratic superpotential in \ref\guk{S.S. Gubser,
I.R. Klebanov, ``Baryons and Domain Walls in an ${\cal N}=1$ Superconformal
Gauge Theory,'' hep-th/9808075.}).

\newsec{Higgs Branches of ${\cal N}=1$ A-D-E Quiver Theories:
The Non-mondromic Case}

Let us recapitulate what these theories are:
Given an affine or ordinary Dynkin diagram of rank $r+1$ or $r$ with
 some labelling of the nodes
$i=0,1,\ldots ,r$ or $i=1,\dots ,r$
we assign a class of field theories with the following
field content. Each node corresponds to an
 ${\cal N}=1$ vector multiplet with gauge
group $U(N_i)$ and an adjoint chiral field $\Phi_i$. Each arrow connecting
the i-th node to the j-th node corresponds to a bifundamental chiral field
$Q_{ij}$ transforming in the fundamental of $U(N_i)$ and antifundamental of
$U(N_j)$. We only consider {\it non-chiral} quivers due to the ${\cal
N}=2$ origen of this theories. This means that
$Q_{ji}$ and $Q_{ij}$ are both present or both absent.

The total gauge group of the theory is then given by $G=\prod_i
U(N_i)$. The tree level superpotential is \superpotential, which we write
here for convinience,
\eqn\supero{
W =\int d^2\theta \sum_{ij} s_{ij}
{\rm Tr}(Q_{ij}\Phi_j Q_{ji}) +\sum_{i} {\rm Tr}P_i(\Phi_i),
 }
where we take
$P_i(x)$ to be a polynomial of degree $s_i+1$, and $s_{ij}$ is the
anti-symmetric form associated to the corresponding Dykin diagram.
In the affine case in addition we have the constraint $\sum_{i=0}^r
 d_i P_i(x)=0$
where $d_i$ are the affine Dynkin indices.
As noted in the
previous section these theories can be geometrically engineered
by fibering wrapped branes around cycles of a fibered A-D-E
geometry over a complex plane.  Moreover, to obtain
this structure we assumed that the fibration does not
induce monodromies on the 2-cycles of A-D-E.  The coupling constant
for the $i$-th gauge group is given by
$$\tau_i=\theta_i+{i\over g_i^2}=B_i^R+{i\over g_s}(r_i^2+{B_i^{NS}}^2)^{1/2}$$
and the $i$-th ${\cal N}=1$ FI terms is given by $r_i$.  Note
that if we set these FI terms to zero, then the coupling constants $\tau_i$
are given by
$$\tau_i =B_i^R+i{B_i^{NS}\over g_s}=B_i^R +B_i^{NS} \tau$$
where $\tau$ is the coupling constant of type IIB strings (here
we also include the possibility of turning on the RR scalar of type IIB
strings as part of $\tau$ ).
Note that in the affine case, due to the fact that $\delta$ has a trivial
$H_2$ class,
$$\sum_{i=0}^{r} d_i \tau_i =0 \quad {\rm mod} \quad (n_1+n_2 \tau )$$
where the mod condition on the right arises because $B$'s are defined
up to addition of an integer.  It is convenient to restrict the domain
of $B$'s such that\foot{Aspects of this will be discussed in detail
in \cfikv .  Note that a particular solution to the above equation is
$\tau_i =d_i/|\Gamma|$ which was identified in \ref\lnv{A.E. Lawrence,
N. Nekrasov and C. Vafa, ``On Conformal Theories in Four Dimensions'',
hep-th/9803015, Nucl. Phys. {\bf B533} (1998) 199.}\ as the one corresponding
to the stringy orbifold ${\bf C}^2/\Gamma$}.
$$\sum_{i=0}^r d_i \tau_i =\tau.$$

Note that these theories can be viewed as deformations of ${\cal N}=2$
supersymmetric gauge theories consisting
of $\prod U(N_i)$, with bifundamental hypermultiplets
according to the edges of the Dynkin diagram.
The deformation to ${\cal N}=1$ is
achieved by adding the superpotential
$W=\sum_{i} \rm{Tr}P_i(\Phi_i)$  (the first term in \supero\
is already induced by the ${\cal N}=2$ supersymmetry).

Now we wish to classify all possible classical solutions of the
field equations.  For simplicity let us consider the case where all the
FI-terms $r_i=0$\foot{We could also consider the case where
all the $B_{i}^{NS}=0$ and $r_i \not=0$ which would also lead to a similar
classification of branches.}.
  For this purpose we should consider the set of
solutions to $dW=0$ modulo complex gauge transformations
\luta .  It turns
out that given the analysis we already did for the trivially fibered
case, one can already read off the answer for the fibered case as well,
as we will now argue.

The field equations obtained from varying the superpotential \supero\
are given by,
\eqn\field{
 \sum_{i} (s_{ij} Q_{ji}Q_{ij}) +
 P'_j(\Phi_j) = 0   \qquad  Q_{ij}\Phi_j = \Phi_{i}Q_{ij}
  \qquad  \Phi_j Q_{ji} = Q_{ji}\Phi_{i},}
for all $j$ and where $P'_k(x)$ is the derivative with
respect to $x$ of $P_k(x)$.  This is essentially the same algebra
we already encountered in the case of the trivially fibered geometry
where now we have $P_j'$ playing the role of $\alpha_j$.  However note
that $P_j'$
commutes will all the elements acting on the j-th vector space. It clearly
commutes with $\Phi_j$.  Moreover from \field\ it follows that $\Phi_j$
and therefore $P'_j(\Phi_j)$ commutes with any chain of the form $Q_{ji}
Q_{ik}\ldots Q_{lj}$
and this gives the totality of operators acting on the $j$-th vector space.
This
implies that $P_j'$ is a c-number in any irreducible representation of this
algebra.  Thus we are back to classifying exactly the same algebra as we
encountered in the trivially fibered case.
We thus borrow the same results we already mentioned:
Let $\rho^k $ be a positive root of the corresponding ADE
Dynkin diagram (ordinary or affine as the
case may be).  For the affine case we also have to choose
how many branches $N$ we choose for the imaginary direction $(0,1)$. Recall
that $\rho^k$, by definition, has an expansion in
simple roots $e^i$ with positive integer coefficients,
$$
\rho^k = \sum_{i} n_i^k e^i.
$$
Moreover all the allowed $\Phi_j$ in a branch have the same
diagonal vev $x$
(due to \field) satisfying
\eqn\bran{
\alpha (\rho^k)=\sum_{i} n_i^k P'_i(x) = 0}
(for the affine case the number $N$ of the branches in the imaginary
direction do not enter as an extra condition
because $\sum_i d_i P_i'$ is identically zero).
This last equation is a polynomial equation and generically has $d_k$
distinct solutions, where $d_k$ is the maximum degree of the $P'_i$'s
involved in the equation. Let $(a,k)$
denote the solutions of \bran\ with $a\in \{1,\ldots d_k\}$.

The gauge group $G$ will generically break as follows,
$$
G=\prod_{i} U(N_i) \rightarrow \prod_{(a,k)}U(M_{(a,k)}),
$$
where $M_{(a,k)}$ denotes the number of times a given branch appears
(with an extra $U(N)$, or its Higgsings, in the affine case
 for the branches which contain
$N$ pure $D3$ branes). This
statement can be made more precise by saying that each time a branch appears
each $\Phi_j$ will have $n^k_j$ eigenvalues equal to the $(a,k)$ solution
of \bran. Therefore, $M_{(a,k)}n_j^k$ is the total number of eigenvalues
equal to the $(a,k)$ solution.  Also the condition that the total
dimension of the i-th vector space is $N_i$ leads to the statement
\eqn\cons{
\sum_{(a,k)}M_{(a,k)}n_i^k = N_i}
(where the left hand side of the above equation has an extra
$Nd_i$ in the affine case).
An example of the above general analysis is
the case considered in \civ\ corresponding to $A_1$. The gauge group is $U(N)$
and the theory has an adjoint field $\Phi$ with superpotential $W(\Phi ) =
P(\Phi )$ where $P(x)$ is a polynomial of degree $k+1$. The classical
equation for the eigenvalues of $\Phi$ is
$P'(x)=g_{k+1}\prod_{i=1}^{r}(x-a_i)$.
Choosing $M_i$ eigenvalues of $\Phi$ to be equal to $a_i$ breaks the gauge
group to $U(M_1)\times \ldots \times U(M_k)$ with $\sum_{i=1}^k M_i= N$.

Here we used a general analysis of the representation of A-D-E quivers.
For the $A_r$ and $D_r$ cases the direct analysis
is simple enough so that the classification can be done very
explicitly using very simple arguments that we will present below.
This analysis turns out to be crucial also for the case of monodromic
fibration, and we will heavily rely on it.
For $E_6$, $E_7$ and $E_8$ it is more complicated and it requires
the use of a
classification theorem of indecomposable representations of quivers proven
in \gab \kac .

Let us show how the branch analysis works explicitly in the $A_r$ and $D_r$
cases.

\subsec{Branches of $A_r$}

The field equations \field\ are given by,
\eqn\An{
Q_{12}Q_{21}+\P{1}{1} = 0 \quad -Q_{21}Q_{12}+Q_{23}Q_{32}+\P{2}{2}=0}
$$
\vdots
$$
$$
 -Q_{r-1,r-2}Q_{r-2,r-1}+Q_{r-1,r}Q_{r,r-1}+\P{r-1}{r-1}=0  \qquad -Q_{r,r-1}Q_{r-1,r}+\P{r}{r}=0,
$$
and
\eqn\commut{
Q_{i,i+1}\Phi_{i+1} = \Phi_{i}Q_{i,i+1} \qquad  \Phi_{i+1} Q_{i+1,i} =
Q_{i+1,i}\Phi_{i} \qquad {\rm for} \qquad i=1,\ldots ,r-1.}
Let us define an operation that we call {\it conjugation by} $Q_{i,i+1}$ as
follows. {\it Given any adjoint field} $\cal F$ {\it of}
$U(N_{i+1})$
 {\it we construct a new adjoint field of } $U(N_i)$ {\it given by}
  $Q_{i,i+1}{\cal
F}Q_{i+1,i}$. A very important case will be
when $\cal F$ is a
polynomial in $\Phi_{i+1}$. In this case, conjugation will have a simple
action, namely, from \commut\ we have
\eqn\conj{
Q_{i,i+1}{\cal F}(\Phi_{i+1})Q_{i+1,i} = {\cal F}(\Phi_i)Q_{i,i+1}Q_{i+1,i}.}

At each node $(j)$ we have three natural adjoint fields, namely, $\Phi_j$,
$Q_{j,j-1}Q_{j-1,j}$ and $Q_{j,j+1}Q_{j+1,j}$. For $j=1$ we define
$Q_{1,0}Q_{0,1}\equiv 0$ and for $j=r$ we define $Q_{r,r+1}Q_{r+1,r}\equiv
0$. The idea is to find a set of three equations for each node that will
only involve adjoint fields at that node. This will provide enough
information about the eigenvalues of each of them since the three adjoint
fields commute among themselves. To see this remember that we showed in the
general discussion that $P'_j(\Phi_j)$ commutes with any chain of
bifundamentals, in particular, with $Q_{j,j-1}Q_{j-1,j}$ and with $Q_{j,j+1}Q_{j+1,j}$.
Finally, considering the field equation,
$-Q_{j,j-1}Q_{j-1,j}+Q_{j,j+1}Q_{j+1,j}+\P{j}{j}=0$ is simple to see that
$Q_{j,j+1}Q_{j+1,j}$ commutes with $Q_{j,j-1}Q_{j-1,j}$.

The first step in getting the equations at the (j)-th node is to conjugate
the first equation in \An\ by $Q_{21}$. Doing so and using \conj\ we get,
$$
Q_{21}Q_{12}(Q_{21}Q_{12}+\P{1}{2}) = 0,
$$
using now the second equation in \conj\ to replace $Q_{21}Q_{12}$ by
$Q_{23}Q_{32}+\P{2}{2}$ we get,
$$
(Q_{23}Q_{32}+\P{2}{2})(Q_{23}Q_{32}+\P{1}{2}+\P{2}{2}) =0.
$$
Conjugating next by $Q_{32}$ and repeating the same procedure until
reaching the $(j-1)$-th node we get,
\eqn\one{
X_j (X_j+\P{j-1}{j})(X_j+\P{j-2}{j}+\P{j-1}{j})\ldots (X_j+\P{1}{j}+ \ldots
+\P{j-1}{j})=0,}
where we have denoted $Q_{j,j-1}Q_{j-1,j}$ by $X_j$ to simplify the equation.

The second equation can be obtained by repeating the same procedure as
before but starting at the last node and conjugating by
$Q_{r-1,r}$. Repeating this until reaching the $(j+1)$-th node will produce
the following equation,
\eqn\two{
Y_j (Y_j+\P{j+1}{j})(Y_j+\P{j+1}{j}+\P{j+2}{j})\ldots (Y_j+\P{j+1}{j}+ \ldots
+\P{r}{j})=0,}
where we have denoted $-Q_{j,j+1}Q_{j+1,j}$ by $Y_j$.

Finally, the last equation we need is the original field equation coming
from the variation w.r.t $\Phi_j$ of the superpotential.
\eqn\three{X_j + Y_j = \P{j}{j}.}

It turns out to be very useful to remember the identification of $P'_i$
with the simple root $\alpha_i$ and introduce the $t_i$ coordinates as in
\labA. This gives $P'_i(\Phi_j) = t_{i}(\Phi_j)-t_{i+1}(\Phi_j)$ for $i=1,
\ldots, r$. In terms of these and shifting $X_j\rightarrow X_j+t_j$ and
$Y_j\rightarrow Y_j - t_{j+1}$ we get field equations of the form,
\eqn\simply{\prod_{i=1}^{j}(X_j+t_{i})=0 \qquad \prod_{i=j+1}^{r+1}(Y_j - t_{i}) =0 \qquad
X_j+Y_j = 0}

Since the three operators commute, we can choose basis in which all three
are diagonal. Let $\vec{v}$ be an eigenvector of $\Phi_j$, $X_j$ and $Y_j$
with eigenvalues $\phi$, $x_j$ and $y_j$ respectively. These three
eigenvalues will have to solve \simply\ when each of the
operators is replaced by its eigenvalue.

The possible solutions are then given by choosing $x_j$ to be $-t_k$ for
some $k=1, \ldots, j$ and $y_j$ to be $t_l$ for some $l=k+1,\ldots
,r$. However, the last equation in \simply\ implies that,
$$ x_j+y_j = - t_{k} + t_{l} = 0$$
or equivalently,
$$ \sum_{m=k}^{l-1}P'_{m}(\phi ) =  0$$
Notice that these are exactly the solutions given in \bran\
since the positive roots for $A_r$ are $\sum_{k=n}^{m}e^k$ with $n\leq
m$.

Finally, we want to show that $\phi$ is an eigenvalue only of $\Phi_m$ for
$m=k,\ldots ,l-1$ and moreover, that the dimension of the corresponding
eigenspaces are all the same.

{}From \commut\ it is easy to see that the $\phi$ eigenspace of $\Phi_i$ is either
sent to the $\phi$ eigenspace of $\Phi_{i+1}$ or belongs to the kernel of
$Q_{i+1,i}$. This implies that we can consider the restriction of the field
equations \An\ to $\phi$ eigenspaces. Finally, we only have to take the
trace of each of the restrictions in \An\ to get,
\eqn\contra{
{\rm Tr}\P{1}{1} + {\rm Tr}\P{2}{2} + \ldots + {\rm Tr}\P{r}{r}=
\sum_{i=1}^r n_i P'_i(\phi ) = 0}
where $n_i$ is the dimension of the $\phi$ eigenspace of $\Phi_i$. But
$\phi$ cannot satisty two polynomial equations simultaneously for generic
$P'_i(\phi )$. Therefore, we are led to conclude that $n_i = 0$ for
$i\not=k,\ldots ,l-1$ and $n_i=n$ for $i=k,\ldots ,l-1$ with $n$ the number
of times we want the given branch to appear.

\subsec{Branches of $D_r$}

Let us use the labels for the nodes given in Figure 1 for $D_r$.
Clearly, the novelty in this case corresponds to the node $(r-2)$. All
branches that do not contain this node are simply $A_{r-3}$ and two
$A_1$'s. For this reason we will only study the field equations
concentrated at the $(r-2)$ node.

The field equations are now given in terms of 3 bifundamentals and one
adjoint field. These are $X= Q_{r-2,r-3}Q_{r-3,r-2}$,
 $Z=Q_{r-2,r-1}Q_{r-1,r-2}$, $Y=Q_{r-2,r}Q_{r,r-2}$
and $\Phi_{r-2}$.

Conjugating from node $(1)$ to node $(r-2)$ we get,
$$X (X + \P{r-3}{r-2})\ldots
(X+\P{r-3}{r-2}+\ldots + \P{1}{r-2}) =0 $$
Conjugating from nodes $(r-1)$ and $(r)$ to node $(r-2)$ we get,
$$ Z(Z+\P{r-1}{r-2})=0  \qquad  Y(Y+\P{r}{r-2})=0 $$
Finally, the equation at node $(r-2)$ is,
$$X + Y + Z = \P{r-2}{r-2}  $$
As for the $A_r$ case, it turns out to be useful to use \labD\ to write
$P'_i = t_i-t_{i+1}$ for $i=1, \ldots , r-1$ and
$P'_r=t_{r-1}+t_r$. The field equations are very simple if we shift,
$X \rightarrow X + t_{r-2}$,
$Y\rightarrow Y - {1\over 2}(t_{r-1}+t_r)$ and $Z\rightarrow
Z-{1\over 2}(t_{r-1}-t_r)$. Then the equations are
given by,
\eqn\Dreq{ \prod_{i=1}^{r-2}(X+t_k)=0 \quad  Y^2={1\over 4}(t_{r-1}+t_r)^2  \quad
Z^2 ={1\over 4}(t_{r-1}-t_r)^2  \quad   X + Y + Z = 0}
In this case, $X$, $Y$ and $Z$ may not commute and we have to be more
careful since some branches might correspond to higher dimensional
representations of these algebra. Let us study each possibility stating
from the one dimensional.

The one dimensional (i.e, $X$, $Y$ and $Z$ are c-numbers) solutions are
easily found. Clearly,
$Y+Z$ can only be $\pm t_{r-1}$ or $\pm t_{r}$. Using
this in the equation for $X$ we get that all the solutions can be collected
in a single equation,
$$
\prod_{i=1}^{r-2}(t^2_{r-1}-t^2_i)(t^2_{r}-t^2_i) = 0
$$
Going back to the language of the simple roots $P'_j$'s we see that the
solutions are in one to one correspondence with the positive roots containing
$P'_{r-2}$ with coefficient one.

Now we can look for two dimensional representations. For this we have to
promote $X$, $Y$ and $Z$ to $2\times 2$ matrices. Notice that if any of
them is proportional to the identity, then using the last equation of \Dreq\
we get that the other two commute and we are back to the one dimensional
case. Therefore, in order to get 2 dimensional representations we need each of
them to have two distinct eigenvalues. For $Y$ and $Z$, notice that this
immediately implies that ${\rm Tr}Y={\rm Tr} Z=0$. This follows from the
fact that the second and third equations in \Dreq\ imply that $Y$ and $Z$
can each only have two eigenvalues that only differ in sign.

Using that $X+Y+Z=0$ we can conclude that ${\rm Tr}X$ is also zero. But in
some basis $X=-\rm{diag}(t_i,t_j)$ for some $i,j\in \{ 1,\ldots r-2\}$ with
$i\neq j$. Therefore the tracelessness condition is simply the following polynomial
equation $t_i(\phi )+t_j(\phi )=0$ where $\phi$ is an eigenvalue of $\Phi_{r-2}$.

Going back to the simple root notation, namely, $P'_i$ we see that
$t_i+t_j$ with $i,j = 1, \ldots ,r-2$ and $i\neq j$ precisely produce all
the positive roots were $P'_{r-2}$ enters with coefficient two.

Using the same argument as for the $A_r$ case, we can prove that for $\phi$
of a given branch associated to a root $\rho^k$
 the corresponding $\Phi_i$ eigenspace will have
dimension $M_k n_i^k$ where $n_i^k$ is given by
$$\rho^k =\sum_j n_j^k e^j$$
and $M_k$ denotes the number of times we choose that branch to appear.

  It is also easy
to show that no irreducible representation with dimension greater
than two exists (similar to the arguments we present in the context
of Laufer's example in section 7.2), and that this analysis completes
the list of all irreducible representations and shows that they are in one
to one correspondence with positive roots of $D_r$.

\newsec{Higgs Branch for Pure D3 Branes}

The case where we only have D3 branes, i.e. where we have $N_i =Nd_i$
for some fixed $N$, gives rise to a theory with $N$ D3 branes.
In particular the case with $N=1$ should have a moduli space given by
the moduli of a D3 brane transverse to the threefold geometry.  This
space should in fact be isomorphic to the space itself.  Let us see
how the moduli space of Higgs branch in this case gives back the
threefold in this case for the affine $A_r$ case.

Consider the case where all $N_i=1$ i.e, the gauge group is $U(1)^{r+1}$.
Let us recall the field equations for the affine $A_r$.
$$ -Q_{0,r}Q_{r,0}+Q_{01}Q_{10}+\P{0}{0} = 0 \qquad
-Q_{10}Q_{01}+Q_{12}Q_{21}+\P{1}{1} = 0$$
$$
-Q_{21}Q_{12}+Q_{23}Q_{32}+\P{2}{2} =0 \quad \ldots \quad
-Q_{r,r-1}Q_{r-1,r}+Q_{r,0}Q_{0,r}+\P{r}{r} = 0$$
In this case $\Phi_i = \phi$ is independent of $i$. Let us use that
 $P'_i=t_i-t_{i+1}$ and recall that we have an extra condition given by
$\sum_{i=0}^{r} P'_i(\phi )=0$ or in
terms of $t_i$'s we get, $t_0 = t_{r+1}$.

Let us define the following coordinates for the moduli space,
$$ u \equiv Q_{0,r}Q_{r,r-1}\ldots Q_{2,1}Q_{1,0} \quad v \equiv
Q_{0,1}Q_{1,2}\ldots Q_{r-1,r}Q_{r,0} \quad \rm{and}\quad  x=Q_{r,0}Q_{0,r} -
t_0$$
The equations of motion imply that we can write $Q_{i-1,i}Q_{i,i-1}=x+t_i$
for $i=0,\dots ,r$.

It is easy to see that $u,v,x,t$ form a complete set of holomorphic
gauge invariant observables.
These three variables are not independent. In order to see the relation, consider,
$$ uv = ( Q_{0,r}Q_{r,r-1}\ldots Q_{2,1}Q_{1,0})  (Q_{0,1}Q_{1,2}\ldots
Q_{r-1,r}Q_{r,0}) =(Q_{0,1}Q_{1,0})\ldots (Q_{r,0}Q_{0,r})$$
where we have used the fact that we are considering abelian gauge theories
in reshuffling the position of the $Q$'s.

Using the field equations we get,
$$ uv = \prod_{i=1}^{r+1}(x + t_i) $$
that is the equation describing the $A_r$ fibration over the $t$-plane \Adef.
Thus we see that the moduli space of the Higgs branch for the D3 brane
is exactly the same as the $A_r$ fibered geometry, as expected. This
is a further confirmation of the picture we have developed.  This
generalizes the same
analysis done in \neg\ for the case of quadratic superpotentials.

\newsec{{\cal N}=1 A-D-E Quiver Theories in the Monodromic Case}

In this section we will begin our discussion of the monodromic
case.  Recall that this is the case where the A-D-E fibration
undergoes Weyl reflections for the 2-cycles of the geometry.
For simplicity we consider the non-affine case.
In general the relevant monodromy will generate a subgroup
of the corresponding Weyl group, which could also be
the full group.  We will discuss examples of all of these cases, but our main
focus will be on the cases corresponding to the monodromy
group being the subgroup
of the full Weyl group generated by Weyl reflections
of all the nodes except for one.  In these cases, as we will discuss,
there is a single blow up mode, corresponding to blowing up the
node with no monodromy.
  This we study in detail,
as it is the simplest case leading
to some new gauge theories, corresponding
to wrapping D5 branes around the blown up $S^2$,
leading to $U(N)$ gauge theory with some matter
content.  It is also the case where quite a bit is known about
the geometry of the local 3-fold \km .

More generally, consider a local Calabi-Yau threefold with
a single $H_2$ element,
represented by a ${\bf P}^1$. An interesting mathematical question
is to classify all such ${\bf P}^1$'s that can shrink inside a Calabi-Yau.
The normal bundle to ${\bf P}^1$
in the blown up CY can be only one of three cases: i) $O(-1)\oplus O(-1)$,
ii) $O(0)\oplus O(-2)$ and iii) $O(1)\oplus O(-3)$.  The case i) is
 rather trivial.
The more interesting cases correspond to ii) and iii).
It turns out that all such ${\bf P}^1$'s can be obtained
by considering monodromic fibrations of A-D-E, where only
one class can be blown up to give rise to ${\bf P}^1$, just as
in the above case. This has
been studied in detail in \km , where case ii) arises
from $A$ quiver theories and case iii) arise in D and E quiver
theories.  In particular case ii) is exemplified
by an A-fibration where the monodromy group is generated
by Weyl reflection about all nodes except for one (any of the nodes).
For case iii) the inequivalent cases can be chosen so that
the special node is the trivalent node of the D and E Dynkin
diagrams--except for one extra $E_8$ case where the special node
is one with Dynkin number 5.\foot{The inequivalent local
geometries can be classified by the Dynkin numbers
of the corresponding A-D-E node, which lead to $A_1$
(1) ,$D_4$ (2),
$E_6$ (3), $E_7$ (4). $E_8$ (5), $E_8$ (6).}

  Now consider wrapping
N D5 branes around such a ${\bf P}^1$.  This gives rise to an
${\cal N}=1$ U(N) gauge theory.  Moreover the number of adjoint matter fields
depend on the number of normal deformations of ${\bf P}^1$ in the Calabi-Yau,
i.e. it should be a holomorphic section of the normal bundle \fibra .  This
implies that in case ii) there
is one and in case iii) there are two.  This matches nicely
the cases one would expect based on asymptotic freedom for the gauge theory.
We find the adjoint matter fields satisfy some constraints.
In some cases, these constraints correspond to critical
points of a superpotential, and  the corresponding
gauge theory can be presented as a  $U(N)$ theory
with one or 2 adjoint fields with some superpotential.

The organization of our discussion for the monodromic case is as follows:
We first present examples of the monodromic A-D-E fibrations.  We then
state the mathematical conditions needed for the fibration to have a single
$S^2$ blow up for cases ii) and iii).
We then analyze the gauge theory of the wrapped branes
around the corresponding $S^2$.
Finally in section 7 we present some detailed examples.

\subsec{Simple Examples of Monodromic Fibration}

 We start
with an A-D-E geometry, and we fiber this over the complex
plane.  However, we consider a situation where the fibration
induces monodromy action on the A-D-E 2-cycles.  In other words,
as we go around various loops in the base of the fibration
the total geometry comes back to itself, but not each individual
2-cycle.  They get reshuffled, according to an element of the
Weyl group of A-D-E.  The monodromy group we obtain will be
a subgroup of the Weyl group.  If the  monodromy
group is the full Weyl group, then this means that there is no
invariant 2-cycle in the fiber geometry, which also implies
that we cannot blow up any 2-cycle in the full 3-fold geometry.
In fact this would be the generic case of A-D-E fibration.
In such a case we will not have any possibility of wrapping
5-branes around 2-cycles of the Calabi-Yau.

However we can also consider situations where the monodromy only
partially mixes the 2-cycles.  In such cases there would be left-over
2-cycles in the full 3-fold geometry for which we can wrap branes around
and obtain non-trivial ${\cal N}=1$ supersymmetric gauge theories.

Let us first give examples of both of these types.  Consider the
$A_1$ case fibered over the $t$ plane:
$$x^2+y^2+(z+t_1(t))(z+t_2(t))=0$$
As noted before, with no loss of generality we can take $t_1=-t_2$.
If $t_1(t)$ is a single valued function of $t$ we are back
to the non-monodromic case we have studied.  However if $t_1$ has
branch cuts as a function of $t$
then we are in the monodromic case.  Suppose for example
$$t_1(t)=-t_2(t)=t^{n+{1\over 2}}$$
In this case as we go around the origin in $t$ plane
we exchange the $t_1\leftrightarrow t_2$, which corresponds
to $A_1$ Weyl reflection.  The geometry in this case is given by
$$x^2+y^2+z^2-t^{2n+1}=0$$
To begin with we have a 2-cycle corresponding to $e_1$ node
of $A_1$. However this cycle does not survive the fibration
because as we go around the origin $e_1\rightarrow -e_1$.  Thus in
the full 3-fold geometry there is no non-trivial 2-cycle.  More precisely
what we mean is that the above  complex geometry
cannot be blown up to yield a 2-cycle.  In such a situation
we cannot wrap D5 branes around any non-trivial 2-cycles as there are none.
One could try to ``force'' a D5 brane in this situation
as follows: Consider $t\not= 0$.  Then there
is a non-trivial 2-cycle over this point, as we just have an $A_1$
geometry.  Wrap the D5 branes around this cycle.  Then as discussed
before this yields a deformed ${\cal N}=2$
gauge theory with superpotential $W(\Phi)\sim {\rm Tr}\Phi^{n+{3\over 2}}$.
However this is only heuristic:  As we mentioned the 2-cycle cannot
be blown up, which means that the quantum 2-volume is zero, so
the gauge coupling is infinite.  The fact that the we are getting
branch cuts in the superpotential is another sign that
forcing a D5 brane in this situation is not allowed.

Let us give one more example along these lines:  Consider
again the $A_1$ case, but now take $t_1=-t_2=\sqrt{t(t-2a)}$
with $a\not=0$. We thus have two points along the $t$ plane
where we get a Weyl reflections $t=0$ and $t=2a$.  Thus again
we would have no non-trivial 2-cycles in the total geometry.
This geometry is given by
$$x^2+y^2+z^2-t(t-2a)=0$$
which by a shift $t\rightarrow t+a$ is equivalent to
$$x^2+y^2+z^2-t^2+a^2=0$$
This is the deformed conifold, and clearly there are no
non-trivial 2-cycles in this geometry as expected.

A more interesting case, from the gauge theory
viewpoint, is when we have a partial monodromy.
Consider for example an $A_2$ geometry fibered over
the plane:
$$x^2+y^2+(z-t_1(t))(z-t_2(t))(z-t_3(t))=0$$
Let us suppose that as we go around $t=0$ the $t_2\leftrightarrow t_3$,
and $t_1$ is invariant.  In other words this generates a $Z_2$
subgroup of the Weyl group of $A_2$ which is given
by permutation group of three elements $S_3$.
 For example let us take the case
$$t_1(t)=b t \qquad t_2(t)=-at +t^{n+{1\over 2}}
\qquad t_3(t)=-at-t^{n+{1\over 2}}$$
This
yields the threefold geometry
$$x^2+y^2+(z-bt)((z+at)^2-t^{2n+1})=0$$
which as expected is single valued over the $t$ plane.
The full threefold geometry does admit a blow up where
we obtain exactly one non-trivial 2-cycle\foot{In this
case only for the case with $n=0$ the blow up geometry is smooth.
 The general condition for the blow up to be smooth will be spelled
out in the next subsection.}.
 To see this, note
that if we identify the nodes $e_1,e_2$ of $A_2$ with $
\alpha (e_1)=(t_1-t_2) $ and $\alpha (e_2)= (t_2-t_3)$
then the Weyl group as we go around $t=0$ acts by
$$e_1\rightarrow e_1+e_2$$
$$e_2\rightarrow -e_2$$
and the combination $ 2e_1+e_2$ is invariant.
So we can blow up this class and wrap
$N$ D5 branes around it and ask what ${\cal N}=1$
gauge theory
it corresponds to?

Note that when we blow up, the quantum volume of $e_2$
is zero, (as $e_2\rightarrow -e_2$ under the Weyl
reflection) so the corresponding gauge coupling
constant is infinite. On the other hand the volume of $e_1$
is finite and is the same as the quantum volume of $e_1+{1\over 2}e_2$.
This implies that the gauge group for the node 1 will be conventional,
whereas for the second node is unconventional.  This suggests
that we should get some insight into this geometry
by concentrating on the gauge theory description
associated to node $1$ in the geometry.  In other words
{\it from the corresponding quiver
theory we only keep the observables which are trivial
under all the other gauge groups, i.e., geometrically
speaking are monodromy
invariant.  This means that we promote certain composite fields
involving chiral multiplets
which are neutral under all the
 would be gauge groups corresponding to nodes
with monodromy and which transform in the adjoint representation of
the gauge group corresponding to the blown up node, as fundamental
fields.}
  We will see
that this approach makes sense, and yields results
which we check with other methods as well.

Let us specialize to a simple example where $A_2
\rightarrow A_1$:
Consider the geometry
\eqn\cygm{x^2+y^2+(z^2+az+t)(z- t)=0}
This corresponds to
$$t_1(t)=t,\qquad t_{2,3}={-a\over 2}\pm {\sqrt{a^2-4t}\over 2}$$
If we consider the ${\cal N}=1$ $A_2$ quiver theory associated to this
case, as studied in section 4, we have two bifundamental fields
$Q_{12}$ and $Q_{21}$ and if we define
$$X=Q_{12}Q_{21}$$
writing the equations we get at the first node, where the
gauge theory has finite coupling constant we have from \one\
and \two\
$$X=P_1'(\Phi_1)=t_1(\Phi_1)-t_2(\Phi_1)$$
$$X(X+P_2'(\Phi_1))=X(X+t_2(\Phi_1)-t_3(\Phi_1))=0$$
Note that $t_1$ is a well defined field as it is single
valued.  However $t_2,t_3$ are not good fields as they
are not single valued functions of $\Phi_1$.  This implies
that $X$ is not a good field.
  To this end we
define
$${\tilde X}=X+t_2$$
in terms of which we obtain
$${\tilde X}=t_1(\Phi_1)= \Phi_1$$
$$({\tilde X}-t_2)({\tilde X} -t_3)=({\tilde X}^2+a {\tilde X}+\Phi_1)=0$$
{}From these equations we see that we can eliminate $\Phi_1$ using
the first equation and obtain
$${\tilde X}^2+{(a+1)}{\tilde X}=0$$
Identifying this as $W'({\tilde X})={\tilde X}^2+(a+1){\tilde X}$,
this can be obtained from a theory with superpotential
$$W({\tilde X})={\rm Tr}\left( {{\tilde X}^3\over 3}+(a+1){{\tilde
X}^2\over 2} \right) $$
Thus we seem to come to the conclusion that we have a single
gauge group with adjoint field ${\tilde X}$ with the above
superpotential. Indeed if we redefine the coordinates
of the CY 3-fold geometry \cygm\ by
$$\rho=z-t-{1\over 2}z(z+a+1)$$
 This corresponds to the geometry
$$x^2+y^2-\rho^2 +{1\over 2}W'(z)^2=0$$
which up to rescalings is exactly the theory corresponding
to a single adjoint field with superpotential $W'$, discussed before.
This gives us confidence that the idea we have proposed for extracting
the gauge theory in situations with non-trivial monodromies is sound.

Before dealing with the general case using the same methods, we turn
to a mathematical discussion of the monodromic fibrations for which
only one node can be blown up.

\subsec{Mathematical Description of Monodromic Fibrations}

Let us describe the threefold geometry in the monodromic case
where the monodromy group is generated
by Weyl reflections about all nodes except for one.  This
discussion parallels that of \km\ but our viewpoint is slightly
different.

We start with a collection of positive simple roots of the A-D-E
geometry, corresponding to a subcollection of the holomorphic spheres.
  These roots generate a subdiagram
$\Gamma$ of the Dynkin diagram.  Monodromy is introduced by
simultaneously taking these spheres to zero size and taking the
subgroup of the Weyl group generated by the associated reflections.
This A-D-E geometry is singular at the points $p_i$ which have
replaced these curves.  These are in one to one correspondence with
the connected components $\Gamma_i$ of $\Gamma$.

Fibering this geometry over the $t$-plane in a generic fashion will
smooth out the singularities, but the threefold geometry can be
singular at the $p_i$ for a non-generic fibering. We will specify
the condition for obtaining a smooth threefold.

The key technical result we use to simplify computations is the
assertion that the geometry near the singularity $p_i$ can be computed
by the following simple procedure: take the equation of the
deformation of $p_i$ (in one of the forms \Adef\Ddef\ given in Section 2), then make a substitution given by
the projection onto the root subspace $V_i$ generated by the component
$\Gamma_i$ of $\Gamma$ corresponding to $p_i$.
A precise statement is given in Theorem~3 of \km.  Instead of setting up
the machinery required for the general case, we illustrate with examples
and refer the reader to \km\ for more details.

\bigskip
\centerline{\epsfxsize=0.75\hsize\epsfbox{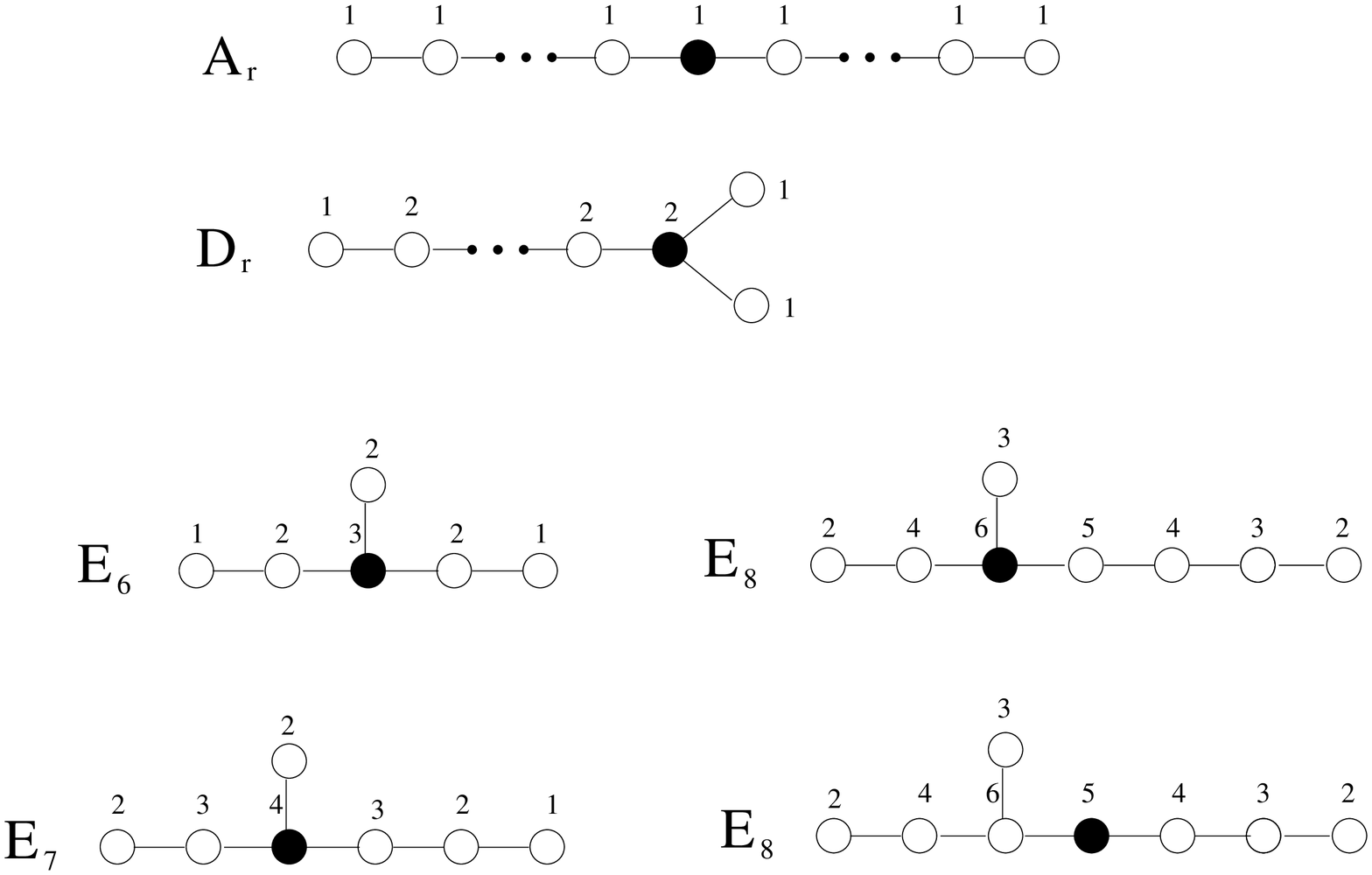}}
\noindent{\ninepoint\sl \baselineskip=8pt {\bf Figure 2}:{\sl
We consider fibrations of A-D-E geometry where the monodromy
group is generated by Weyl reflection about all nodes
except for the black node.  }}
\bigskip

{\bf $A_n \rightarrow A_1$ via a partial resolution }

We illustrate with the case of an $A_n$ singularity where only one vertex
is blown up.

Recall the general deformation of an $A_n$ for which all
$n$ vertices of the Dynkin diagram get blown up.
\eqn\adown{ x^2+y^2+\prod_{i=1}^{n+1}(z+t_i)=0  \qquad
\sum_{i=1}^{n+1}t_i = 0}
If we only want to
blow up vertex number $k$, we are left with a surface $Y_0$ with a single
$P^1$ with two singularities, an $A_{k-1}$ and an $A_{n-k}$
(we modify conventions accordingly if $k=1$ or $k=n$; in those
cases there is only one singularity).  Denote by $Z_0$, the blown up geometry
of
$n$ curves in an $A_n$ configuration.   As discussed
in section 2, the space of complex deformations
${\rm Def}(Z_0)$ is the hyperplane $\sum t_i=0$ in the affine space with
coordinates $t_1,\ldots, t_{n+1}$, the corresponding deformation being
described by \adown.

The deformation space ${\rm Def}(Y_0)$ is a quotient of ${\rm Def}(Z_0)$ by
$S_k\times S_{n-k+1}$, i.e. the Weyl group generated by all nodes except
for the k-th node.
 The coordinates on ${\rm Def}(Z_0)$ are naturally given
by appropriate symmetric functions of the $t_j$ :

$$\sigma_i(t_1,\ldots,t_k),\ i=1,\ldots,k,\
\st_i(t_{k+1},\ldots,t_{n+1}),\ i=1,\ldots,n-k+1.$$

where
$$\sigma_i(x_1,\ldots ,x_n)=\sum_{j_1,j_2,...,j_i\ {\rm distinct}}
x_{j_1}...x_{j_i}.$$
Only $n$ of these are independent coordinates on ${\rm Def}(Z_0)$ due to the
relation

$$\sigma_1(t_1,\ldots,t_k)+\st_1(t_{k+1},\ldots,t_{n+1})=0.$$

The equation \adown\ becomes

\eqn\awdown{x^2+y^2+(z^k+\sum_i\sigma_iz^{k-i})(z^{n-k+1}+\sum_i\st_iz^{n-k-i+1})
=0.}

The CY geometry is obtained by letting the $\sigma_i,\st_i$ be
holomorphic functions of $t$.  If they all vanish at $t=0$
(which we will not in general require) it was shown in \km\ that
the condition for the single blow up to be smooth near $t=0$ is that $\sigma_k(t)$
and
$\st_{n-k+1}^2(t)$ vanish to first order in $t$.

\bigskip

{\bf $D_r\to \; $trivalent vertex}

If we blow down all but the trivalent vertex of the $D_r$, we get an
$A_{r-3}\times A_1\times A_1$ subdiagram with Weyl subgroup
$S_{r-2}\times {\bf Z}_2\times {\bf Z}_2$.  With reference to the
coordinate system $t_1,\ldots,t_r$ given in
equation \Ddef , the $S_{r-2}$ acts on the first
$r-2$ coordinates, and the ${\bf Z}_2$ reflections are
$$\eqalign{(t_{r-1},t_r)&\mapsto(t_r,t_{r-1})\cr
(t_{r-1},t_r)&\mapsto(-t_r,-t_{r-1})\cr}.$$
The invariant coordinates are $\sigma_i=\sigma_i(t_1,\ldots,t_{r-2})$
as well as $\tilde\sigma_1^2=(t_{r-1}+t_r)^2$ and $\tilde\sigma_2=t_{r-1}t_r$.

The projections of \km\ Theorem~3 can be described in terms of two projections,
first onto the space $V_1$ spanned by the first $r-2$ roots and second onto the
space $V_2$ spanned by the last 2 roots.  The first describes an $A_{r-2}$
singularity, and the second gives the $A_1\times A_1$ (which we conveniently
think of as a $D_2$).  Near the
$A_r$ the geometry is given by
$$x^2+y^2+z^{r-1}+\sum_{i=1}^{r-1}\sigma_i(t)z^{r-i}=0.$$
This is smooth at $(x,y,z,t)$ unless $x=y=0$ and
$z^{r-1}+\sum_{i=1}^{r-1}\sigma_i(t)z^{r-i}=0$ has a multiple root at $(z,t)$.
Near the $D_2$, we find the local equation by putting $r=2$ in \Ddef\
and rewriting in terms of the invariant coordinates.  We get
$$x^2+y^2z
+{(z-t_{r-1}^2)(z-t_r^2)-t_{r-1}^2t_r^2 \over z}+2 t_{r-1}t_r y=0$$
or
\eqn\geoDtwo{
x^2+y^2z+z+(2\tilde\sigma_2(t)-\tilde\sigma_1^2(t))+2\tilde\sigma_2(t)y=0.}
The geometry \geoDtwo\ is smooth at $(x,y,z,t)=(0,0,0,0)$ when
$\tilde\sigma_1^2(t)$ and $(\tilde\sigma_1^2-4\tilde\sigma_2)(t)$ have
simple zeros at $t=0$.  Thus, assuming that the $t_i$ vanish
at $t=0$ (which we will not necessarily assume) then the condition
to have a smooth blow up is that $\tilde\sigma_1^2(t)$ and
$(\tilde\sigma_1^2-4\tilde\sigma_2)(t)$ have
simple zeros at $t=0$

If desired, the $D_2$ can be replaced by two local calculations near each
of the $A_1$s.  The result is of course the same.

\bigskip

{\bf $E_n\to\;$trivalent vertex}

We now consider the $E_n$ case with $n=6,7,8$ where the blown up
sphere corresponds to the trivalent vertex.
Suppose we blow down all holomorphic spheres except the one corresponding
to the trivalent vertex of $E_n$.  Then $\Gamma=A_2\times A_{n-4}\times A_1$.

We present the roots and consequently the Weyl group action as in \km.
For convenience of the reader, the $W(A_{n-1})=S_n$ subgroup acts on
$(t_1,\ldots,t_n)$ in the usual way, while the reflection in the remaining
root is given by sending $(t_1,\ldots,t_n)$ to
%
\eqn\reflect{
\left(
t_1-{2\over3}T,
t_2-{2\over3}T,
\ldots,
t_3-{2\over3}T,
t_4+{1\over3}T,
\ldots,
t_n+{1\over3}T
\right)
}
where $T=t_1+t_2+t_3$.

The $W(A_2\times A_{n-4}\times
A_1)$ invariant coordinates are

$$\sigma_j=\sigma_j\left(t_1-{T\over 3},
t_2-{T\over 3},t_3-{T\over 3}\right),\qquad j=2,3$$
$$\tilde\sigma_j=\sigma_j\left(t_4+{T\over6},\ldots ,
t_n+{T\over6}\right)\qquad j=1,\ldots,n-3$$
$$\rho_2=T^2.$$

Near the $A_2$ singularity, the local geometry is given by
$$x^2+y^2+z^3+\sigma_2(t)z+\sigma_3(t)=0.$$
Near the $A_{n-4}$ singularity, the local geometry is given by
$$x^2+y^2+z^{n-3}+\tilde\sigma_1(t)z^{n-4}+\ldots+\tilde\sigma_{n-3}(t)=0.$$
Near the $A_1$ singularity, the equation is
$$x^2+y^2+z^2+\rho_2(t)=0.$$

So the condition for
smoothness at $z=t=0$ is that the expressions
$\sigma_3,\ \tilde\sigma_{n-3}$, and $\rho_2$
all have simple zeros at $t=0$.   General values of $(z,t)$ are similar.

%
%
%
%
%
%
%

\bigskip

{\bf Second $E_8$ case}

Suppose we blow down all holomorphic spheres except the one corresponding
to the vertex of $E_8$ with Dynkin number 5.  Then $\Gamma=A_4\times A_3$.
We have already described the Weyl group action in the last section.

The first four $W(A_4\times A_3)$ invariant coordinates are

$$\sigma_j=\sigma_j(s_1,s_2,s_3,s_4,s_5),\qquad j=2,3,4,5,$$
where
$$\eqalign{
s_1&=t_1-{2\over5}\left(t_1+t_2+t_3+t_4\right)\cr
s_2&=t_2-{2\over5}\left(t_1+t_2+t_3+t_4\right)\cr
s_3&=t_3-{2\over5}\left(t_1+t_2+t_3+t_4\right)\cr
s_4&=t_4-{2\over5}\left(t_1+t_2+t_3+t_4\right)\cr
s_5&={3\over 5}(t_5+t_6+t_7+t_8)}.$$
The remaining four coordinates are
$$\tilde\sigma_j=\sigma_j\left(t_5+{1\over5}{\tilde T},t_6+{1\over5}{\tilde
T},
t_7+{1\over5}{\tilde T},t_8+{1\over5}{\tilde T}\right)\qquad j=1,\ldots,4$$
where ${\tilde T}= t_1+t_2+t_3+t_4$.
The geometry near the $A_4$ is given by
$$x^2+y^2+z^5+\sigma_2(t)z^3+\sigma_3(t)z^2+\sigma_4(t)z+\sigma_5(t)=0$$
and near the $A_3$ it is
$$x^2+y^2+z^4+\tilde\sigma_1(t)z^3+\ldots+\tilde\sigma_4(t)=0.$$

So the condition for
smoothness at $z=t=0$ is that the expressions
$\sigma_5$ and $\tilde\sigma_4$
have simple zeros at $t=0$.   General values of $(z,t)$ are similar.

\subsec{Gauge theory description}

We would now like to describe the gauge theory corresponding
to D5 branes wrapping the blown up $S^2$ in the
A-D-E fibered geometries described
above.  We will consider all cases, except the last $E_8$
case with the special node corresponding to Dynkin number 5.

The strategy to find the gauge theories obtained when we wrap $N$ D5 branes on
these 2-cycles is to consider the corresponding quiver theory for the full
A-D-E, but restrict attention to the fields which are charged only under
the $U(N)$ gauge group associated to blown up node, and neutral
under all other groups. This in particular entails promoting certain composite
fields to fundamental fields of the $U(N)$ gauge theory.  Algebraically
the description of the neutral fields and the equations they satisfy
parallels the mathematical description given above, and this is related
to the fact that both involve monodromic invariant data.

Let us start with $A_r$ quiver theories with the special node
corresponding to the $k$-th node as in Figure 2.  Then $Q_{k,k-1}
Q_{k-1,k}$ and $Q_{k,k+1}Q_{k+1,k}$ and $\Phi$ (the
adjoint field corresponding to the $k$-th node) generate
all the fields invariant with respect to other gauge groups,
as can be verified from the equations discussed in section 4.  Moreover,
as discussed in section 4, denoting a (a suitably shifted)
$Q_{k,k-1}Q_{k-1,k}$,$Q_{k,k+1}Q_{k+1,k}$ by $X$, $Y$ respectively,
we find that they satisfy some particular
 field equations given by \simply. Let us rewrite \simply\
 (changing the sign in the definition of $Y$),
$$\prod_{i=1}^{k}(X+t_{i})=0 \qquad \prod_{j=k+1}^{r+1}(Y+t_{j}) =0 \qquad
X - Y = 0 $$
where $t_i$ should be viewed as functions of the field $\Phi$
given by $t_i(\Phi)$.
These can now be written as,
$$ X^{k} +\sum_i \s_{i}X^{k-i} =0  \qquad
Y^{r-k+1}+\sum_{i}\st_{i}Y^{r-k+i+1} =0 \qquad  X-Y=0$$
where $\s_i$, $\st_i$, X, Y are invariant under the Weyl
reflections we are interested in.  This is very similar
to the equation describing the geometry \awdown.
The first two equations, substituted by
$X=Y$ from the third equation and identifying
$X=Y=z$, appended with $x=y=0$ of equation \awdown\ give  precisely the
condition for
singular points in the geometry.  Solutions to these
equations are precisely the positions where we can have a blown up
${\bf P}^1$ from the geometry analysis.  From the gauge theory analysis
this is very natural as this is also classifying the possible choices
of deformation of the gauge theory, by giving vev to $X=Y$, and should
correspond to supersymmetric vacua, which are in 1-1 correspondence
with allowed ${\bf P}^1$'s.\foot{The extra condition $x=y=0$ from \awdown\
can be viewed, perhaps,
as a trivial addition of two additional massive adjoint fields
to the gauge theory, which have been integrated out.}

Consider now the $D_r$ case. In this case, as can be
seen from the analysis of section 4, the fields
generating the relevant chiral ring is generated by the adjoint
field $\Phi$ and the
composite adjoint
fields $X,Y,Z$ (suitably
shifted with functions of $\Phi$) corresponding to the three edges
adjacent to the trivalent
node.  We wish to write the chiral
ring relations for these fields.
 These equation were also worked out in section 4 \Dreq.
  Let us rewrite them here,
$$ \prod_{i=1}^{r-2}(X+t_k)=0 \quad  Y^2={1\over 4}(t_{r-1}+t_r)^2  \quad
Z^2 ={1\over 4}(t_{r-1}-t_r)^2  \quad   X + Y + Z = 0$$
These can now be written as,
\eqn\Drmonogen{\sum_{i=0}^{r-2}\sigma_i X^{r-2-i}=0 \quad  Y^2={1\over 4}(\tilde{\sigma}_1)^2  \quad
Z^2 ={1\over 4}(\tilde{\sigma}_1)^2-\tilde{\sigma}_2  \quad   X + Y + Z = 0 }
where $\s_i$, $\st_i$, $X$, $Y$, and $Z$ are invariant under the Weyl
reflections.

Finally, let us consider the $E_r$ cases with the special
node being the trivalent vertex.  Again the chiral
fields are generated by the three relevant meson fields together
with $\Phi$.
 These cases were not studied in
section 4 therefore we will have to start from the original form of the
equations at the trivalent node.
 We will use the label for the nodes given in Figure 1.

The field equations at the trivalent node are given by,
$$ X(X+P'_2)(X+P'_2+P'_1)=0 \qquad Y\prod_{i=4}^{r}(Y+P'_4+\ldots + P'_i) =0 $$
$$ Z(Z+P'_r) = 0   \qquad  X+Y+Z=P'_3$$
Shifting $X\rightarrow X-{1\over 3}(P'_1+2P'_2)$, $Z\rightarrow Z-{1\over
2} P'_r$ and $Y\rightarrow Y+{1\over 6}(3P'_r+2P'_1+4P'_2+6P'_3)$ and using
the coordinates given in \labE\ we can write the equations in terms of
invariant fields as follows,
$$X^3+\s_2 X+\s_3 = 0 \qquad  Z^2 - {1\over 4}\rho_2 = 0 \qquad
Y^{r-3}+\st_1 Y^{r-4}+\ldots + \st_{r-3} =0  \qquad X+Y+Z = 0$$
where $\s_i, \st_i$ and $\rho_2$ turn out to be exactly equal to those
defined in the mathematical analysis for $E_r$ at the trivalent node.

Let us now try to identify the field theories that will give rise to these
equations. We will study first the examples with trivalent nodes. The $A_r$ case will be
studied in full detail as the first example in the next section.

We have seen that for all $D_r$ and $E_r$ cases,
 the field equations in terms of invariant
fields can be written as follows,
$$X^{r+1}+\s_{1}X^{r}+\ldots + \s_{r+1} = 0  \qquad Y^{q+1}+\sh_{1}Y^{q}+\ldots + \sh_{q+1} = 0$$
\eqn\gian{Z^{p+1}+\st_{1}Z^{p}+\ldots + \st_{p+1} = 0   \qquad   X+Y+Z = 0}
where $p,q,r$ denote the total number of extra nodes on each
of the three legs of the Dynkin diagram
emanating from the trivalent vertex.
Note that in addition
the coefficients of the constraints are functions of the adjoint
field $\Phi$.  Furthermore, the quiver analysis given in section
4 implies that, in addition to the above equations we have to impose
\eqn\comm{[X,\Phi]=[Y,\Phi]=[Z,\Phi]=0}

Note that the constraints satisfied by the fields parallels
the mathematical description of the geometry.  As noted above
this is not surprising as both parameterize the position of the blow up
sphere.
The local geometry discussion we reviewed earlier in the
paper, provided three local
patch descriptions, which we naturally identify with the three
polynomial constraints above.  However, the geometry description
of \km\ does not provide a global description  of the blown up geometry
in terms of these patches.  The gauge theory construction above
suggests, however, that there must exist a
simple global description of the blown
up geometry where the three local coordinate patches get identified
with $X,Y,Z$ which in turn should be identified with three sections of $O(1)$
over the ${\bf P}^1$.  Moreover the fact that they satisfy a linear
condition such as $X+Y+Z=0$ is natural given that there
are only two independent holomorphic sections for
$O(1)$.  We will verify that the geometry does satisfy this description
at least in some cases that we present as examples in the next section.

It is natural to ask whether all
these constraints arise from a physical gauge system
with a superpotential.  The answer
to this question appears to be in the negative, except for
a very special class, that we will identify.
The explanation
of this is as follows:  In the superpotential
the mixing between $X,Y,Z$ and $\Phi$ is fixed by the
original quiver theory to be
$$W={\rm Tr} (X+Y+Z)\Phi +...$$
which suggests that we could interpret $X+Y+Z=0$ equation
as the $dW/d\Phi=0$ equation.  If so, this would imply
that the other three equations should be viewed as $dW/dX,dW/dY,dW/dZ$
respectively.  However, this is possible only if
$${\partial^2W \over \partial \Phi \partial X} =
{\partial^2W \over \partial X  \partial \Phi } \qquad {\rm etc.}$$
which implies that the equation involving the polynomial in $X$ should
have only an additive linear dependence in $\Phi$, and the same
goes for the polynomial equations in $Y$ and $Z$.  In other
words $\s_{r+1} ,\sh_{q+1},\st_{p+1}$ are linear in $\Phi$.
and all the other $\sigma$'s are constants.  Note that in this case
the condition of commutativity \comm\ is {\it automatic}.  In particular
if we commute any of the polynomial equations with the corresponding field
we get the corresponding commutativity condition \comm .  This is as
it had to be if we were to be able to find a superpotential giving
the constraints, because we cannot obtain 7 equations by considering
gradient of a superpotential involving only 4 fields.  There should have
been three redundant equations\foot{Actually
there is always one redundant equation.  The commutativity
of $X$ and $\Phi$ follows from the other
two commutativity conditions, given the constraint $X+Y+Z=0$.}.  It is not
difficult to see that this
condition alone would essentially force us to the above structure on the
form of the $\Phi$ dependence of $\sigma$'s for the compatibility of the above
constraints with the existence of the superpotential. It is interesting to note that
the condition of linearity of the highest indexed $\s$ with respect
to $\Phi$ near the blow up point
was exactly the condition for the blow up to be smooth,
as discussed in the previous subsection.

So now let us determine
the superpotential for the case where all $\s_i$, $\st_i$ and $\sh_i$
are constants except for
$\s_{r+1}$, $\sh_{q+1}$ and $\st_{p+1}$ that are taken to be linear
functions of $\Phi$. This implies that the field equations can be written
as,
\eqn\laex{R'_{r+2}(X) = \Phi \qquad  Q'_{q+2}(Y) = \Phi  \qquad  P'_{p+2}(Z)=\Phi
\qquad {\rm and } \qquad X+Y+Z=0  }
where $R_{r+2}$, $Q_{q+2}$ and $P_{p+2}$ are polynomials of degree $r+2$,
$q+2$ and $p+2$ respectively.

Now it is clear that this equations can be obtained from a superpotential
of the form,
$$ W = {\rm Tr}\left( R_{r+2}(X) + Q_{q+2}(Y) + P_{p+2}(Z) -\Phi (X+Y+Z)\right)  $$
(In the A case we obtain the same except we set $Z=0$ in the above).

Now we can write an effective superpotential in terms of two adjoint fields
by integrating out $\Phi$ and $Z$. This results in,
\eqn\sula{ W(X,Y) ={\rm Tr}\left( R_{r+2}(X) + Q_{q+2}(Y) + P_{p+2}(-X-Y)
\right) }

In the next section we will explore in detail some examples that illustrate
the general relations found in this section.


\newsec{Examples of Monodromic Fibrations}

In this section we provide several examples where the gauge theory
obtained from wrapping a D5 brane around the $P^1$ that survives the monodromy
induced in the geometry can be described very explicitly.  As we saw, the
gauge theory description suggests a rather simple global geometric
description of the blown up ${\bf P}^1$ for all cases. However such
a mathematical construction is not currently known in the generality
suggested by the gauge theory.  Instead only some explicit blown up
geometries
are known in detail and we shall discuss them and see that they
match the gauge theory description precisely.

 First, we will
study some examples of monodromic $A_r$ fibrations whose geometry is
globally equal to that of an $A_1$ case, where the blown up node is
the first node of the $A_r$.
 The last two examples are related to monodromic
$D_r$ fibrations. We first analyze the geometry introduced by Laufer
\laufer\ as an example of a
$P^1$ with normal bundle $O(1)\oplus O(-3)$ and then we generalize to all
$D_r$ monodromic fibrations.  The mathematical discussion is rather involved,
but we present it for the sake of completeness.  We also specialize
the gauge theory discussion of the previous section to these cases
and see a perfect match between the geometric description and the
gauge theoretic description.

\subsec{ $A_r \rightarrow A_1$ }

Let us start with the gauge theory construction. Let us assume
that the monodromic $A_r$ fibration is such that the geometry
dictates a Weyl reflection on $A_{r-1}\subset A_r$, where
we identify the nodes of $A_{r-1}$ with the nodes $2,\ldots ,r$.

Using the general analysis of the previous section we can write the
equations at the first node as,
\eqn\twoeq{X= t_1  \qquad  \prod_{i=2}^{r+1} (X-t_i) = \sum_{j=0}^{r} (-1)^j
\sigma_j X^{r-j}=0}
where $\sigma_j$ are the symmetric functions of $t_2,\ldots , t_{r+1}$.

We will consider the special case where $\s_i$ are constants for
$i=1,\ldots ,r-1$ except for $\sigma_r$ and $t_1$ to be linear in $\Phi$
plus a constant.  Let us take $t_1=\Phi$ and $\sigma_r=-\alpha \Phi$.
{}From the first equation in \twoeq\ we get $X=\Phi$ and using
this we can eliminate $\Phi$ from the second equation in \twoeq\ to yield
$$\sum_{j=0}^{r-1} (-1)^j \sigma_j
X^{r-j}-\alpha X=0$$
which we interpret as $W'(X)=0$, i.e. the critical points
of the superpotential $W(X)$.

The CY geometry for the case where $\s_i$ are constant can be shown to be
globally the same as an $A_1$ geometry. Let us now show that
explicitly. The geometry is given in general by,
$$ z^2+y^2 + (\sum_{j=0}^{r} (-1)^j \sigma_j x^{r-j})(x - t_1)=0 $$

Writing the geometry in an expanded form
$$  z^2+y^2 + (x^r+\sigma_1 x^{r-1}+\ldots + \sigma_{r-2}x^2+(\sigma_{r-1}-\alpha)x+
\alpha (x - t))(x - t) =0 $$
Let $\omega =(x-t)$ and $W'(x) =x^r+\sigma_1 x^{r-1}+\ldots +
\sigma_{r-2}x^2+(\sigma_{r-1}-\alpha)x$.
Then,
$$ z^2+y^2 + (W'(x)+ \alpha \omega )\omega =0$$
by letting $\rho=\omega+{1\over 2\alpha}W'(x)$ we get the $A_1$ geometry expected
for this superpotential
$$ z^2+y^2+ \rho^2 - {1\over \alpha^2} W'^2(x) =0.$$

This shows that our identification of the field theory was correct because
D5-branes wrapping the non trivial two cycles of this geometry has already
been engineered in \kkal\ and studied in detail in \civ.  Note that the more general form of the
dependence of $\sigma$'s on $\Phi$ will not yield a threefold geometry
in the above form, further strengthening the argument that in these
cases the gauge theory with a single adjoint field cannot be described
by a superpotential term.

\subsec{Laufer's Example: the D-case}

Consider the local geometry described in \laufer. This is the classical example of an exceptional
$P^1$ with normal bundle $O(1)\oplus O(-3)$. We will see that the blown
down geometry corresponds to a monodromic $D_r$ fibration where only the
trivalent node can be blown up and corresponds to the $P^1$. The deformation
theory can be worked out explicitly and a superpotential can be
obtained. We study the ${\cal N}=1$ gauge theory with two
adjoint fields obtained by wrapping
D5-branes in the $P^1$ and work out the analysis of critical points of the
superpotential. The field equations give an algebra that admits one and two
dimensional representations (as we expect from the fibering
of $D$ geometries), and we find a prefect match
between these representations and the degree
one and two curves in the 3-fold geometry.

\bigskip
{\bf Resolved Geometry}
\bigskip

In this section we present a detailed mathematical description
of a blown up ${\bf P}^1 $ with normal bundle being $O(1)\oplus O(-3)$.

We glue together two copies of ${\bf C}^3$.  The first ${\bf C}^3$ has
 coordinates $(x,y_1,y_2)$ and
the second has coordinates $(w,z_1,z_2)$.  The glueing data is
$$\eqalign{
z_1&=x^3y_1+y_2^2+x^2y_2^{2n+1}\cr
z_2&=x^{-1}y_2\cr
w&=x^{-1}.}$$
Consider the curve $C$ given in the first
patch by $y_1=y_2=0$ and in the second patch by $z_1=z_2=0$, with $x$ and
$w$ being the coordinates along $C$.  The curve $C$ has normal bundle
$O(1)\oplus O(-3)$ as can be seen from the terms in the glueing data which
are linear in the equations of $C$.

The deformation theory of the curve has already been worked out in
\potential.  If we consider the deformation of the equation
$y_2=0$ for $C$ to $y_2=ax+b$ and deform $y_1=0$ to $y_1=
-((ax+b)^{2n+1}-b^{2n+1})/x$, we obtain
$$z_1=b^2+2abx+(a^2+b^{2n+1})x^2,\qquad z_2=a+bx^{-1}.$$
Note that the deformation parameters $a,b$ can be identified
with vev of massless adjoint scalars if we consider wrapping
$D5$ branes around the curve $C$.
This deformation of $z_2=0$ is automatically holomorphic in $w=x^{-1}$ (and
in fact fixed the general choice of deformation of $y_2=0$).  The
deformation of $z_1=0$ is holomorphic in $w$ precisely when
$2ab=a^2+b^{2n+1}=0$.  The condition for having a holomorphic
deformation should translate to the condition of satisfying
critical points of a superpotential.  In fact,
  these equations are precisely the critical points of
the holomorphic function
\eqn\pot{W(a,b)=a^2b+{b^{2n+2}\over 2n+2}}
which we therefore identify with a superpotential (including
 a trace in front, if we consider more than one wrapped brane).

The only solution of \pot\ is $a=b=0$, but this is a degenerate solution.
We deform to a generic situation by adding terms to \pot\ modulo its
partial derivatives.
%
The deformed geometry is conveniently written as
$$\eqalign{
z_1&=x^3y_1+y_2^2+\alpha x +x^2(y_2P(y_2^2)+Q(y_2^2))\cr
z_2&=x^{-1}y_2\cr
w&=x^{-1}.}$$
where $P$ has degree exactly $n$ and $Q$ has degree at most $n$.

This geometry is blown down by four holomorphic functions.
$$\eqalign{
v_1&=z_1=x^3y_1+y_2^2+\alpha x +x^2(y_2P(y_2^2)+Q(y_2^2))\cr
v_2&=w^2z_1-z_2^2-\alpha w=xy_1+y_2P(y_2^2)+Q(y_2^2)\cr
v_3&=wv_2-z_2P(z_1)-wQ(z_1)\cr
v_4&=z_2v_2-wz_1P(z_1)-z_2Q(z_1)}$$
The last two equations are linear in $z_2$ and $w$, and can be solved as
\eqn\vars{z_2={-v_3v_1P(v_1)-v_2v_4+v_4Q(v_1)\over
v_1P(v_1)^2-(v_2-Q(v_1)^2)},
\qquad w={-v_4P(v_1)-v_2v_3+v_3Q(v_1)\over v_1P(v_1)^2-(v_2-Q(v_1)^2)}.}
We substitute \vars\ and $v_1=z_1$ into the equation for $v_2$ and obtain
\eqn\deqn{
v_2(v_1P(v_1)^2-(v_2-Q(v_1))^2)=v_4^2-v_1v_3^2+\alpha(v_4P(v_1)+v_2v_3-v_3
Q(v_1)).}

This is the blown-down geometry.
The degree 1 curves are obtained by solving
$$\eqalign{
2xy+\alpha&=0\cr
x^2+yP(y^2)+Q(y^2)&=0.}$$
which leads to the superpotential
\eqn\supsh{W=yx^2+R(y)+\alpha x}
where $R'(y)=yP(y^2)+Q(y^2)$.
This superpotential has been obtained from the geometry
\potential.
Suppose that we have a solution $(x,y)=(-\alpha/2d,d)$ for $d\ne0$.
There are generically $2n+3$ such solutions.
  If $d=0$,
then $\alpha=0$, $x=0$, and $Q$ has no constant term, which implies from
the equations that $\alpha=O(d)$.  In any case, we get the
equation
\eqn\disc{\alpha^2+4y^2(yP(y^2)+Q(y^2)).}

The deformed curve is

$$\eqalign{
y_1&=\dots\cr
y_2&=d-{\alpha\over 2d}x}$$
where we do not need the exact form of $y_1$, only needing to observe that
its form suffices to guarantee that the equation $z_1=0$ deforms
holomorphically.  Substituting into the $v_i$ gives for the location of the
singular curves
$$\eqalign{
v_1&=d^2\cr
v_2&={\alpha^2\over 4d^2}\cr
v_3&={\alpha\over d}P(d)\cr
v_4&={\alpha^3\over 8d^3}+{\alpha\over 2d}Q(d)}$$

In this case and all subsequent cases, it can be checked directly from the
blown-down equation that these singular points are conifolds for generic
deformations.

The degree 2 curve has equations $y_1=y_2^2=0$ in the first patch, and
$z_1=z_2^2=0$ in the second patch.  Now for the deformed degree 2 curves,
we let $c$ be a root of $P$ .  There are $n$ such choices.
  The deformed curve is at
$$\eqalign{
y_1&=\ldots\cr
y_2^2&=c-\alpha x-Q(c)x^2.}$$
If $\alpha=Q(c)=0$, one might naively think that this has split up into two
disjoint curves $y_2=\pm c^{1/2}$, but they meet in the other patch so are
a connected curve. Substituting into the $v_i$ we get
$$\eqalign{
v_1&=c\cr
v_2&=Q(c)\cr
v_3&=0\cr
v_4&=0}$$

The variables here should be matched to \bkl\ or equivalently to \small.  A partial match identifies
$x_1$ of \bkl\ with our $P$, and the defining equation of $D_1$ in \bkl\
with our \disc.

The geometry given in \deqn\ can be rewritten as a $D_r$ fibration with
monodromies. In order to see this let us introduce the following change of
variables.
Let $x=v_4+{1\over 2}\alpha P(v_1)$, $y=v_3$, $z=v_1$ and
$t=v_2-Q(v_1)$. Then \deqn\ can be written as,
\eqn\Dmono{x^2-zy^2+(t+Q(z))(t^2-zP^2(z))
-{1\over 4}\alpha^2 P^2(z)+\alpha t y = 0}
This will lead us to an alternative derivation of the above
result for the superpotential based on the quiver analysis
we have discussed in this paper.

We have seen above that this
 geometry has $2n+3$ isolated non-trivial degree 1 curves
and $n$ isolated degree 2 curves.  What this should imply
in the corresponding gauge theory
is that if we consider a $U(N)$ version of this theory,
the inequivalent choices of Higgs branch will have $2n+3$ inequivalent
one dimensional representations and $n$ two dimensional representations.
We will verify this claim in the next section, which is rather non-trivial
for the relation of degree two curves and irreducible representations
of the $dW=0$ equations.

\bigskip

{\bf The Monodromic Quiver Construction and Laufer's Example}

\bigskip

We will construct now a gauge theory that will correspond to the geometry
in \Dmono . The idea is to use a particular monodromic $D_r$ fibration. In
the previous section we studied the most general case. So we only have to
borrow the field equations \Drmonogen\ in terms of invariant fields,
\eqn\feDr{\sum_{i=0}^{r-2}\sigma_i X^{r-2-i}=0 \quad  Y^2={1\over 4}(\tilde{\sigma}_1)^2  \quad
Z^2 ={1\over 4}(\tilde{\sigma}_1)^2-\tilde{\sigma}_2  \quad   X + Y + Z =
0.}
%




%

In order to induce the monodromy we want we have to impose that $\sigma_{r-2}$,
$\tilde{\sigma}_1^2$ and $\tilde{\sigma}_2$ be linear
functions of $\Phi$. The rest of them will be taken to be constants
independent of $\Phi$.

Let $\sigma_{r-2}=\Phi + b$, $\tilde{\sigma}_1^2= 4 \Phi$ and
$\tilde{\sigma}_2= \alpha $. The equations are now given by,
$$Y^2=\Phi  \qquad Z^2 +\alpha = \Phi  $$
$$  -\sum_{i=0}^{r-3}\sigma_i X^{r-2-i}
-b = \Phi \qquad X+Y+Z =0,$$
which we have written it in the same form as presented in \laex .  Using
this we can immediately borrow the results from the previous section and
write the superpotential from \sula\ which gives
$$W={\rm Tr} \left( {1\over 3}Y^3-(\sum_{i=0}^{r-3}{1\over r-1-i}\sigma_i X^{r-1-i}
+b X)-{1\over 3}(X+Y)^3- \alpha (X+Y) \right).$$
Note that for the choice of coefficients leading to the Laufer
geometry the $Y^3$ piece above cancels out.  After a shift
$Y\rightarrow Y-{1\over 2}X$ and removing an irrelevant overall minus sign we obtain
$$
W(X,Y) = \rm{Tr}\left( XY^2 + R(X) + \alpha Y \right) .
$$
where $R'(X) =\sum_{i=0}^{r-3}\sigma_i X^{r-2-i} + {1\over 4}X^2
+ b + {1\over 2}\alpha$. Note that this is identical with the superpotential arrived at
by geometric reasoning leading to \supsh\ , with the change
in notation ($x\leftrightarrow Y$, $y\leftrightarrow X$) and the
definition of $P$ and $Q$ by the identification
$R'(X) =  XP(X^2)+Q(X^2)$.  This gives
$$P(z)=z^n+\s_2 z^{n-1}+...+\s_{2n-2} z +\s_{2n}$$
\eqn\defquo{Q(z)=\s_1 z^n +\s_3 z^{n-1}+...+(\s_{2n-1}+{1\over 4})z+
(b+{1\over 2} \alpha ),}
where we have assumed odd $r$ (to obtain Laufer's geometry)
with $r=2n+3$.
 Now that we have succeeded in reproducing
the superpotential we have to show that the geometry of
this fibration
matches \Dmono.

\bigskip

{\bf Recovering the blown-down geometry}

We now have discussed which monodromic $D_r$ quiver
theory gives rise to the superpotential expected for Laufer's
example. Here we show that this threefold is indeed
the one corresponding to Laufer's example given by \Dmono .

Recall that the $D_r$ fibration geometry is given by \Ddef,
\eqn\geoD{ x^2+y^2z
+{\prod_{i=1}^{r}(z-t_i^2)+(-1)^r\prod_{i=1}^rt_i^2 \over z}+2 \prod_{i=1}^r
t_i
y }
Notice that we have rescaled $t_k$ by a factor of $i$ as well as $y$ and
$z$ in order to get the above equation.

Now we want to write this in terms of the invariant coordinates in the
deformation space. For this notice that,
\eqn\pro{\prod_{i=1}^{r-2}(z-t_i^2) = \left\{ \eqalign{
& (z^n+\s_2 z^{n-1}+\ldots +\s_{2n})^2-z(\s_1
z^{n-1}+\ldots + \s_{2n-1})^2  \quad  \rm{for} \quad  r =2n+2  \cr &
z(z^n+\s_2 z^{n-1}+\ldots +\s_{2n})^2-(\s_1 z^n+\ldots + \s_{2n+1})^2 \quad
\rm{for} \quad r =2n+3}
\right.}
It turns out that Laufer's geometry is reproduced if we take the odd case
$(r=2n+3)$.
We also need the remaining two factors in product that appears in \geoD,
$$ (z-t^2_{2n+2})(z-t^2_{2n+3}) = z^2+( 2\st_2 - \st_1^2 )z+\st^2_2$$
The gauge theory fibration tells us that we simply have to replace $\Phi$
by $t$ to get the new geometry. This implies that $\st^2_1 = 4 t$ , $\st_2 =\alpha$ and $\s_{2n+1}=t+b$.

Using all this and the two polynomials $P(z)$ and $Q(z)$ defined in
\defquo\ we get that the geometry \geoD\ can be written as,
\eqn\variab{\eqalign{& x^2+ zy^2 + 2\alpha (t+b)y + \left(
zP^2(z)-(Q(z)-{1\over 4}z-{1\over 2}\alpha +t)^2
\right) \cr
& + {\alpha\over
z}\left( (zP^2(z)-(Q(z)-{1\over 4}z-{1\over 2}\alpha +t)^2)(2z+\alpha
)+\alpha (t+b)^2
\right) = 0}}
Now we only have to use some change of variables. Let us
shift $t\rightarrow t - Q(z)+{1\over 4}z+{1\over 2}\alpha$ and $y\rightarrow
y-{\alpha\over z}(Q(z)-{1\over 2}\alpha-b-{1\over 4}z)$. Notice that the shift
in $y$ is well defined for $z=0$ since $Q(z)-{1\over 2}\alpha-b$ has at least a
simple zero at $z=0$. Then \variab\ is given by,
$$ -4 (zP^2(z)-t^2)(t-Q(z))+\alpha^2 P^2(z)+z y^2-2 \alpha ty+x^2 = 0 $$
Finally, we only need to rescale $y\rightarrow -2y$, $x\rightarrow 2ix$ and
$t\rightarrow -t$ to get,
$$ x^2 -zy^2+(t+Q(z))(t^2-zP^2(z))-{1\over 4}\alpha^2 P^2(z)+\alpha ty =0 $$
This is in perfect agreement with \Dmono.

\bigskip
{\bf Physical Analysis of Critical Points of the Superpotential}

The gauge theory that is proposed above is an
${\cal N}=1$ theory with two chiral superfield $X$ and $Y$.
We will now analyze the choices of the branches of this theory, by
finding the critical points of $W$ and check that they
beautifully match the structure
of holomorphic curves of degree 1 and 2 expected in the Laufer's geometry.

The superpotential is
given by,
$$
W(X,Y) = {\rm Tr}\left( XY^2 + R(X) +
\alpha Y \right).
$$
The field equations are,
\eqn\anti{XY+YX=-\alpha  \quad \rm{and} \quad Y^2 + R'(X)=Y^2+XP(X^2)+Q(X^2) =0}
We want to find all possible irreducible
representations of this algebra. For this one
has to realize that $X^2$ and $Y^2$ are Casimirs. This is easy to see since
the first equation in \anti\ implies that $[X^2,Y]=0$ and $[Y^2,X]=0$.

Let $X^2=x^2 \id$ and $Y^2 = y^2\id$. Then the second equation in \anti\
can be written as,
\eqn\dimen{y^2 \id + P(x^2)X+Q(x^2) = 0.}

Clearly, if $P(x^2) \neq 0$ then $X$ is a c-number and this implies that
$Y$ is also a c-number. These are the 1-dimensional representations. To see
how many of then are there, let us multiply \dimen\ by $x^2$, and use $2 xy =
\lambda$ to write it as follows,
$$ P(x^2)x^3 + Q(x^2)x^2 + {1\over 4}\lambda^2 = 0$$
This equation has $2n+3$ solutions and that is the number of 1-dimensional representations.

Two dimensional representations, if any, should correspond to
$P(x^2)=0$.  There are $n$ such solutions.  Choose any of these solutions
which correspond to a fixed $x^2$.
This also fixes $y^2$ from \dimen .  From the first equation in \anti\
one can shift $X\rightarrow X+a Y$ and $Y\rightarrow Y+bX$ such that
$X^2=Y^2=0$ and $XY+YX=c$.  There is only one irreducible
representation of this algebra, and that is a two dimensional
representation corresponding to the fock space for the realization
of a single fermionic
creation/annihilation algebra.

Note that the solutions for one (two) dimensional representations
match the location of the degree one (two) curves in the geometry
as they should.

\subsec{Most general $D_r$ geometry}

Laufer's geometry is not the most general Calabi-Yau with a blow up $P^1$
corresponding to a trivalent node of a $D_r$ monodromic fibration.
As we have seen, Laufer's
example is (essentially) the only such $D_r$ case which admits a
superpotential description.  However, we could consider more general
monodromic $D_r$ fibrations which do not admit a superpotential description.
We would like to check that also in these cases the geometry description
matches the gauge theory analysis based on quivers.

We will first display the mathematical computation for
the most general $D_4$ for simplicity and
then show how the result generalizes to $D_r$ for any $r$.

\bigskip

{\bf General Geometric Construction of Monodromic $D_r$ }

We want to construct a Calabi-Yau threefold as a deformation of the partial
blowup of a $D_4$ singularity where only the curve corresponding to the
trivalent vertex of the Dynkin diagram has been blown up.

We recall the general deformation of the minimal resolution of the
$D_4$ singularity (see \km\ for more details
and references).  The deformation parameters are $t_1,\ldots,t_4$, and the
general deformation is a blowup of
\eqn\gend{x^2+{(yz+\sh_4)^2-\big((z^2-\sh_2z+\sh_4)^2+z(-\sh_1z+\sh_3)^2\big)\over z}=0.}
Here the $\sh_i$ are elementary symmetric functions of $t_1,\ldots,t_4$.
In fact, the full Weyl group of $D_4$ acts on the $t_i$ in the usual way.

The blowup is achieved in 2 steps.  We recall the first step completely
but only partially do the second step since we do not need the full blowup.
The equation can be rewritten as
$$(x+\sh_1z-\sh_3)(x-\sh_1z+\sh_3)+(y-z+\sh_2)(yz+z^2-\sh_2z+2\sh_4)=0.$$
Now blow up the ideal $J=(x+\sh_1z-\sh_3,y-z+\sh_2)$.\foot{There are 12 branches
of the singular locus, lying over the respective loci $t_i=\pm t_j$ in
$t$-space.  These are in a natural one to one correspondence with the
positive roots of $D_4$.
The ideal $J$ vanishes on the part of
the singular locus $t_3+t_4=0,y=t_1t_2,z=-t_3^2$ corresponding to one of
the exterior vertices of
the Dynkin diagram.}  The interesting part of the blowup is in the patch
$U=(x+\sh_1z-\sh_3)/(y-z+\sh_2)$.  The equation becomes after eliminating $x$
\eqn\firstblow{\eqalign{
&U\big(U(y-z+\sh_2)-2\sh_1z+2\sh_3\big)+yz+z^2-\sh_2z+2\sh_4\cr
=&(y-z+\sh_2)(z+U^2)+2U(-\sh_1z+\sh_3)+2(z^2-\sh_2z+\sh_4)\cr
=&\big(y-z+\sh_2+2(z-\sh_2-\sh_1U-U^2)\big)(z+U^2)+2(U^4+\sh_1U^3+\sh_2U^2+\sh_3U+\sh_4)}.}
This is visibly a deformed $A_3$.  We can complete the blowup by blowing up
the $A_3$ as in \km\ to get the general
deformation of the fully resolved $D_4$.

As we are only interested in
blowing up the middle vertex of the $A_3$ (which corresponds to the central vertex of the $D_4$), we only blow up partially, and introduce a new set of
variables.
We let $\s_i$ be the elementary symmetric functions of $t_1,t_2$
and $\st_i$ be the elementary symmetric functions of $t_3,t_4$.
This corresponds to the change of variables
$$\eqalign{
\sh_1\rightarrow&\s_1+\st_1 \cr
\sh_2\rightarrow& \s_2+\st_2+\s_1\st_1\cr
\sh_3\rightarrow& \s_1\st_2+\s_2\st_1\cr
\sh_4\rightarrow& \s_2\st_2,\cr}
$$
or said differently, to the factorization
$$U^4+\sh_1U^3+\sh_2U^2+\sh_3U+\sh_4=(U^2+\s_1U+\s_2)(U^2+\st_1U+\st_2).$$
The middle vertex is blown up by blowing up the ideal
$(y-z+\sh_2+2(z-\sh_2-\sh_1U-U^2),U^2+\s_1U+\s_2)$.  So we introduce
homogeneous coordinates $(r,s)$ on a $P^1$ and write
$$s\big(y-z+\sh_2+2(z-\sh_2-\sh_1U-U^2)\big)=r(U^2+\s_1U+\s_2).$$
There are two patches, obtained by putting $s=1$ and $r=1$ respectively.
In the first patch, we get
\eqn\rpatch{
\eqalign{
\big(y-z+\sh_2+2(z-\sh_2-\sh_1U-U^2)\big)&=r(U^2+\s_1U+\s_2)\cr
r(z+U^2)+2(U^2+\st_1U+\st_2)&=0,
}}
where the second equation in \rpatch\ is obtained by substituting the
first equation into \firstblow.  Note that the first equation of \rpatch\
can be used to eliminate $y$, so that the second patch can be completely
described by the second equation of \rpatch, obtaining a hypersurface in
the variables $r,z,U,t_i$.

Since the exceptional set of the first blowup is not desired, we blow
down its proper transform obtained after the second blowup.  This
exceptional set is located at $\st_1=0,\ z=\st_2$; it then follows
from \rpatch\ that $r=-2$.  We blow this down by the change of
variables
$$a=\st_1+(r+2)U,$$ which leads after a little algebra to\foot{The
form of this equation could have been deduced immediately from the
results of \small.  We
can use the identity $r(r+2)=(r+1)^2-1$ and a change of variables to
realize this as a local $D_2$ deformation.}

\eqn\partblowdown{a^2+\big(r(r+2)\big)z+2\st_2(r+2)-(\st_1)^2=0.}
A nice check is that $\st_1$ only appears to an even power.  This was
required because the reflection
$$(t_1,t_2,t_3,t_4)\mapsto(t_1,t_2,-t_4,-t_3)$$
in the Weyl group required for the blowdown
takes $\st_1$ to $-\st_1$.  In fact, $\s_1,\s_2,\st_2,(\st_1)^2$
generate the ring of invariants of the $Z_2^3$ subgroup of $W(D_4)$
generated by the 3 reflections in the 3 exterior roots, and may be used as
coordinates on the space of deformations of the partially blown up $D_4$.
These are essentially the same as the coordinates used in \km\ and
\bkl, differing by a linear change
of variables.

The blowdown is realized by
\eqn\blowdown{\eqalign
{
x&=U(y-z+\sh_2)-\sh_1z+\sh_3\cr
 &=-\sh_1z+\sh_3-z(a-\st_1)+\big(a\s_1+(r+2)\s_2+\s_1\st_1\big)\cdot
{a-\st_1\over r+2}\cr
&=-\sh_1z+\sh_3-z(a-\st_1)-\s_1(rz+2\st_2)+\s_2(a-\st_1)\cr
y&=z-\sh_2-2(z-\sh_2-\sh_1U-U^2)+r(U^2+\s_1U+\s_2)\cr
 &=-z+\sh_2+(a-\st_1)\s_1-rz-2\st_2+\s_2r\cr
 &=-(r+1)z+a\s_1+(r+1)\s_2-\st_2\cr
z&=z
}}

Similarly, in the second patch we get the equations
\eqn\spatch{
\eqalign{
s\big(y-z+\sh_2+2(z-\sh_2-\sh_1U-U^2)\big)&=U^2+\s_1U+\s_2\cr
z+U^2+2s(U^2+\st_1U+\st_2)&=0.
}}

We could similarly calculate the blowdown in the $s$ patch, but we will see
that we can get away with setting $t=0$ in \spatch.  Setting $b=(2s+1)U$
we get the equation

\eqn\pres{(s+1)b^2=s(2s+1)y.}

The blown-down $r$-patch \partblowdown\ and corresponding blown-down
$s$-patch glue together to give a space ${\cal X}$ describing the
general deformation of the partially blown-up $D_4$.

We can now at last construct the most general Calabi-Yau threefold
geometry of this type.  We express the deformation parameters
$\s_1,\s_2,\st_2,(\st_1)^2$ as holomorphic functions of a
parameter $t$ defined in a neighborhood of $t=0$.  With this
substitution, \gend\ describes a singular hypersuface $Y$ in a space
with coordinates $(x,y,z,t)$, exhibited as deformation of $D_4$.  The
desired Calabi-Yau threefold $X$ is obtained from ${\cal X}$ by
performing the same substitution for $\s_1,\ldots,(\st_1)^2$ in terms of $t$.
The blowdown map $f:X\to Y$ is computed by making the same substitution in
\blowdown.  We now focus attention on the curve $C$ which is blown down by
$f$.

The exceptional curve $C$ in the blown-down $r$-patch is given by $a=z=t=0$,
and in the blow-down $s$-patch by $b=y=t=0$.

We now compute the deformation space (Hilbert scheme) of $C\subset X$,
 starting with first
order deformations of $C$, which are described by
Hom$(I/I^2,{\cal O}/I)$, where $I$ is the ideal sheaf of $C$ in $X$.
It is straightforward to check from
\partblowdown\ and \pres\ that
$z$ and $y$ are torsion classes modulo $I^2$, so must map to 0 under a
homomorphism corresponding to a first order deformation.  Furthermore,
from the equation $b=a/r$, we conclude that
the space of first order deformations is
\eqn\firstorder{\eqalign{
a&\mapsto \e_1+\e_2(r+1)\cr
z&\mapsto 0\cr
b&\mapsto \e_1s+\e_2(1+s)\cr
y&\mapsto 0\cr
}}
where the quadratic terms in $\e_1,\e_2$ are set to 0 so that we are describing
a first order deformation.  As a check, \partblowdown\ is satisfied after
the first two substitutions of \firstorder, and \pres\ is
satisfied after
the last two substitutions of \firstorder.
We have shifted $r$ to $r+1$
for convenience in subsequent computations.
We identify
the $\e_i$ with the values of the 2 chiral fields, and compute deformations
to deduce the critical points of the superpotential.

We deform $C$ by expressing $a,z,t$ as holomorphic functions of $\e_1,\e_2,r$
subject to the constraint \partblowdown.  Since deformations of $C$ must blow
down via \blowdown, we see that $x,y,z,t$ must be independent of the local
coordinate $r$ on $C$.  In particular $z=z(\e),\ t=t(\e)$.

Now express $a$ as a power series in $(r+1)$ and substitute into \partblowdown.
We see that that $(r+1)$ can only occur linearly.  So redefining the local
coordinates $\e_1$ and $\e_2$ if necessary, we can assume without loss of
generality that $a=a(\e,r)$ is still given by the first equation in
\firstorder .

Now expanding \partblowdown\ as power series in $(r+1)$ and collecting
terms, we get
\eqn\constraint{\eqalign{
\e_1^2-z(\e)+2\st_2(t(\e))-(\st_1)^2(t(\e))&=0\cr
\e_1\e_2+\st_2(t(\e))&=0\cr
\e_2^2+z(\e)&=0\cr
}}

These are the conditions that the deformation of the affine part of $C$
in the $r$-patch stays in $X$.  The conditions that the curve
blows down under \blowdown\ comes from the $y$ and $x$ equations:
\eqn\cblows{\eqalign{
-z+\e_2\s_1+\s_2&=0\cr
-z\e_2-\s_1z+\s_2\e_2&=0\cr}}
where we have suppressed the $t(\e)$ dependence of some of the terms.
If \constraint\ and \cblows\ are both satisfied, then the compact curve
$C$ deforms without having to check in the $s$ patch, since the fibers of the
blowdown map, of which \blowdown\ is an affine piece, are compact.
{}From the last equation of \constraint\ we get $z=-\e_2^2$.  Then \constraint\
and \cblows\ reduce to the three equations
\eqn\modulicorr{\eqalign{
\e_1^2+\e_2^2+2\st_2-(\st_1)^2&=0\cr
\e_1\e_2+\st_2&=0\cr
\e_2^2+\e_2\s_1+\s_2&=0\cr}}

The equations \modulicorr\ are interpreted as the equations of the natural
correspondence $(C_\e,t)$ consisting of pairs of the deformed curve $C_\e$
and $t$ such that $C_\e$ is contained in the deformation of the partial
blowup of $D_4$ corresponding to parameter $t$.  So the deformation space
(Hilbert scheme) of $C$
in $X$
is locally defined by eliminating $t$.\foot{ Eliminating $\e_1,\e_2$ yields
the discriminant.  This corrects an error in \bkl. There it was claimed
that the Hilbert scheme was defined by the discriminant.  We see from the
above that the correspondence projects to the Hilbert scheme, and to the scheme
in the disc defined by the discriminant under the other projection.  In
particular, if the Hilbert scheme is discrete, then its multiplicity coincides
with the multiplicity of the discriminant, as claimed in \bkl.}

Now we generalize the result \modulicorr\ to $D_r$ with $r=2n+3$.
 This is achieved by
simply replacing the last of the three equations by a polynomial of degree
$2n+1$ in $\e_2$. The result can then be given by,
\eqn\modulicorrgen{\eqalign{
\e_1^2+\e_2^2+2\st_2-(\st_1)^2&=0\cr
\e_1\e_2+\st_2&=0\cr
\e_2^{2n+1}+\s_1\e_2^{2n}+\ldots +\s_{2n+1}&=0\cr}}

These equations are the ones which we compare with the gauge theory
analysis in the next section.

\bigskip
{\bf Comparison to Gauge Theory}

The gauge theory obtained from the geometry just discussed when we wrap
$D_5$ branes is as in Laufer's example a $D_r$ quiver theory with
monodromies.

The equations at the trivalent node were obtained in \feDr\ and are given by,
\eqn\whl{\sum_{i=0}^{r-2}\sigma_i X^{r-2-i}=0 \quad  Y^2={1\over 4}(\tilde{\sigma}_1)^2  \quad
Z^2 ={1\over 4}(\tilde{\sigma}_1)^2-\tilde{\sigma}_2  \quad   X + Y + Z = 0}

We want to show that these equations are equivalent to \modulicorrgen. For
this we only have to notice that $X+Y+Z=0$ implies that we can parametrize
those three fields in term of two new fields $\e_1$ and $\e_2$. Let us
define the following parametrization,
\eqn\mapDr{\eqalign{Y &={1\over 2}(\e_1-\e_2) \cr
                    Z &=-{1\over 2}(\e_1+\e_2) \cr
                    X &=\e_2\cr}. }
With this the first three equations in \whl\ can easily be seen to be
identical to \modulicorrgen. This proves that the structure of one dimensional
representations agree with that of the geometry.  The higher dimensional
representations arise by promoting these to $N\times N$ fields, and it would
be interesting to verify that they match the geometric description for
enumeration of higher
degree curves.

This success encourages us to propose that also for $E_r$ cases the
gauge theory result based on monodromic quivers holds.  It would be interesting
to construct the corresponding blown up geometries and check that the
deformation spaces match the physical prediction given in \gian .

\bigskip

\newsec{Large N duals of ${\cal N}=1$ A-D-E quiver theories}

For a generic ${\cal N}=1$ A-D-E quiver theory, each branch
is equivalent in the IR to a pure ${\it N}=1$ gauge theory
involving a product of some unitary groups.  As such one expects
to have gaugino condensation in each gauge group.  For a single $U(N)$ theory, it
was proposed \kls\nm\va\ that the large $N$
description leads to a geometry were the $S^2$ has shrunk and $S^3$ has
grown together with some flux.

For the more general A-D-E quiver theory we thus also expect a similar
thing to happen.
The geometries studied in section 3 become singular after blowing down the
$S^2$'s. In order to recover a smooth space one can deform the complex
structure and this will generate non-trivial 3-cycles with the topology of
$S^3$, which should have the interpretation of appropriate gaugino
condensation.

The simplest example $A_1$ was studied in \civ\ where the equation of the
singular space was given in \conif,
$$
x^2+y^2+z^2+W'^2(t) = 0.
$$
In that case it was argued that the most general deformation of the
geometry was,
$$
x^2+y^2+z^2 + W'^2(t)+ g_{p-1}(t) = 0,
$$
where $W$ is a polynomial of degree $p$ in $t$ and
where $g_{p-1}(t)$ is a polynomial of degree $p-1$. These deformations
depend on $p$ parameters, which get identified with the choice of the
$p$ parameters denoting the branches of the theory (i.e. as to how
we distribute the branes among various vacua).  This map was found
by extremizing the superpotential in this dual geometry where
$$W=\int H\wedge \Omega$$
for suitable $H$ flux through the 3-cycles.
  It is natural
to ask how one can generalize these transitions to the general
case of A-D-E quiver theories under study.
For simplicity we first consider non-monodromic A-D-E quiver theories,
and discuss the generalization to monodromic cases in the context
of Laufer's example.  Moreover in the context of non-monodromic A-D-E
theories, we first comment on the
ordinary A-D-E case and at the end discuss the generalization to the
affine case, which involves only a minor modification of the discussion.

For simplicity we consider the case where the superpotentials
$W_i(\Phi_i)$ are all polynomials of degree $p+1$.  All the other
cases can be viewed as deformations of this case by adjusting
coefficients of the superpotentials.
As discussed in section 4, for the case where the
superpotential is a polynomial of degree $p+1$ in all the
adjoint fields, we expect to have $pR_+$ branches, where
$R_+$ is the number of positive roots:
$$2R_++r=dim(G)$$
This corresponds to the fact that we can break the $\prod U(N_i)$
to $\prod U(M_\alpha)$ where $\alpha$ runs over $pR_+$ branches
and $M_\alpha$ denotes how many of those branches we have.

We should now show that desingularization of the $W$ will
allow exactly as many coefficients as these branches, and
as many $S^3$'s.  The latter fact is simple to establish:
There are exactly as many shrunk $S^2$'s as the branches.
That is how we found the branches, namely by the condition
of a vanishing $S^2$. Thus each one of them can be deformed
to an $S^3$.   We will now show that there are also exactly
as many normalizable deformations of the geometry as the
number of branches.
One can easily deduce, from
our discussion in section 2 and the relation of the
gradient of superpotential to $\alpha_i$, that for degree $p+1$
superpotentials that we are considering the 3-fold geometry
is given by
$$f=f_{A-D-E}(x,y,z)+a t^{pc_2}+ ...=0$$
where $c_2$ is the dual coxeter number of the corresponding
A-D-E, and we have indicated
explicitly the term with the highest power in $t$.
  The number of normalizable deformations of this geometry
(including the log normalizable ones) is given by
\gvw\shapv,
$$D={1\over 2} dim {\cal R}+{1\over 2}d_{\hat c/2}$$
where $dim{\cal R}$ denotes the dimension of the singularity
ring
$${\cal R}: \qquad {\bf C}[x,y,z,t]/[dW=0]$$
and $d_{\hat c/2}$ is the number of fields of charge $\hat c/2$,
where we assign charge $1$ to $W$ and charge to the variables
$x,y,z,t$ compatible with the highest power having charge $1$,
and $\hat c$ is the highest charge of the ring ${\cal R}$.

 From the above equation, one sees that $dim{\cal R}=(r)(pc_2-1)$.
This follows from the fact that A-D-E singularity corresponding
to rank $r$ gauge group has
$r$ ring elements, and $t^{pc_2}$ has $pc_2-1$ ring elements.
Also $d_{\hat c/2}=r$ because for each element of the A-D-E
ring there is a unique monomial $t^\alpha$ which if we multiply it
with, makes it
have charge $\hat c/2$.  Thus we have
$$D={1\over 2}r(pc_2-1)+{r\over 2}={rpc_2\over 2}$$
We now use the identity
$$dim G=(1+c_2)r$$
to write
$$R_+={1\over 2}(dim G-r)={1\over 2}(r c_2)$$
which shows that $pR_+=rpc_2/2=D$ which is indeed the number
of inequivalent branches we have.  The story from this
point onward is identical to the analysis of \civ.  Namely
we have $D$ resolved conifolds which can have the topology of
$S^3$.  These are in one to one correspondence with $\gamma =(a,k)$
where $k$ runs over positive roots and $a$ runs from $1$ to $p$.
If we have $M_\gamma$ branes in the $\gamma$ branch, this corresponds
to turning $M_\gamma$ units of RR flux through the corresponding
$S^3$.  Moreover we have to turn on the dual B-fluxes to correspond
to the coupling constant of the corresponding gauge field.  Moreover the
strength of the corresponding flux for branch $\gamma$ is given by
$$H_{B-cycles}=\rho^k\cdot (\sum_{i=1}^r \tau_i e_i)$$
where $\tau_i={i\over g_i^2}+\theta_i$
denotes the coupling constant  of the gauge group
corresponding to the $e_i$ node of A-D-E,
and can be identified with the dual $B=B_R+\tau B_{NS}$ through
the $i$-th $S^2$ cycle.  Again, as in \civ\
one expects the running of the coupling constant and this
introduces an IR cutoff in geometry, playing the role of UV cutoff
in gauge theory.

We thus obtain the superpotential
$W=\int H\wedge \Omega$ \tv\mayr\
in terms of $M_\gamma$
which when extremized determines
the coefficients of the deformed geometry.
Dynamical aspects of these theories as well as
the relation of Seiberg-like dualities to Weyl reflections of
A-D-E (and affine A-D-E)
will appear in \cfikv.

As for the affine case the story is very similar to what
we discussed above.  The main difference is that here we have
an additional integrally indexed label for the choice of branches, related
to the number of D3 branes.  Even though this does
affect the geometry by giving rise to the corresponding
 $5$-form field strength in the internal geometry, it does not
 affect the complex geometry and the
superpotential, which depends only on the $H$ fluxes.
Thus for each branch of affine A-D-E, we consider its projection
to the A-D-E, which is labeled by a root of A-D-E $\pm \rho^k$
and a choice of an integer from $1,...,p$ for each such root.
The main difference from the previous case, as far as the superpotential
is concerned is that now we obtain not only the positive roots of A-D-E
but also the negative roots.  This simply translates to the statement
that the corresponding fluxes can be positive or negative and we
only have to include the net flux.  The rest
of the discussion is parallel to the A-D-E case above.  Thus as far
as the total possibilities of superpotentials for various branches,
we get for the affine versus
the non-affine case exactly the same, except
that what plays the role of the rank of the gauge group in the
non-affine case, can be positive or negative in the affine case.

\subsec{Extension to Large $N$ Dualities for Monodromic Quiver Theories}

The same ideas should work in the more general case
of monodromic A-D-E fibration.  For concreteness
let us discuss the case of Laufer's example studied in this paper.
In this case the large $N$ dual should correspond to deforming
the complex geometry given by \Dmono .  Recall that the Higgs
branch was characterized by the splitting of the
branes in one dimensional representations for which there are $2n+3$
inequivalent choices, and two dimensional representations for which
there are $n$ inequivalent choices.  Thus
we expect to have $(2n+3)+n=
3n+3$ normalizable (or log normalizable) deformations for \Dmono .
Setting $P(z)=z^n, Q(z)=\alpha =0$ as an example, we see that
\Dmono\ reduces to
$$x^2-zy^2+t^3-t z^{2n+1}=0$$
We wish to find the number of normalizable deformations
of this geometry.  This is given by the number of elements of the
singularity ring, with charge less than or equal to half the maximum
value in the ring.
One can
readily check that this is $3n+3$ as expected.
Thus these $3n+3$ parameters can be fixed
by extremizing the superpotential as in other cases.
This provides new result for gauge theories, where one
can obtain exact information about the quantum corrected
superpotential for an ${\cal N}=1$ $U(N)$ gauge theory
with two adjoint superfields, with the above superpotential
(i.e. the deformation of D-type superpotential).  Similarly
one can analyze other E-type cases as well.

\centerline{\bf Acknowledgements}
We would like to thank A. Adem, M. Aganagic, J. Bryan, M. Douglas, B. Fiol,
K. Intriligator, and N.C. Leung
for valuable discussions.

The research of FC and CV is supported in part by NSF grants PHY-9802709
and DMS-0074329.
The research of SK is supported in part by NSF grant DMS-073657 and NSA
grant MDA904-00-1-052.
\listrefs
\end